\documentclass[aps,prl,reprint,superscriptaddress,twocolumn,longbibliography]{revtex4-1}
\usepackage{amsfonts,amssymb,amscd,amsthm}
\usepackage{graphicx}
\usepackage{mathrsfs}
\usepackage{soul,color,xcolor}
\usepackage[intlimits]{amsmath}
\usepackage[colorlinks, citecolor=red]{hyperref}
\usepackage{etoolbox}
\usepackage{upgreek}
\usepackage[T1]{fontenc}

\begin{document}
\title{Long-time storage of a decoherence-free subspace logical qubit \\ in a dual-type quantum memory}

\author{Y.-L. Xu}
\thanks{These authors contribute equally to this work}%
\affiliation{Center for Quantum Information, Institute for Interdisciplinary Information Sciences, Tsinghua University, Beijing 100084, PR China}

\author{L. Zhang}
\thanks{These authors contribute equally to this work}%
\affiliation{Center for Quantum Information, Institute for Interdisciplinary Information Sciences, Tsinghua University, Beijing 100084, PR China}

\author{C. Zhang}
\thanks{These authors contribute equally to this work}%
\affiliation{HYQ Co., Ltd., Beijing 100176, P. R. China}

\author{Y.-K. Wu}

\affiliation{Center for Quantum Information, Institute for Interdisciplinary Information Sciences, Tsinghua University, Beijing 100084, PR China}
\affiliation{Hefei National Laboratory, Hefei 230088, PR China}

\author{Y.-Y. Chen}
\affiliation{Center for Quantum Information, Institute for Interdisciplinary Information Sciences, Tsinghua University, Beijing 100084, PR China}

\author{C.-X. Huang}
\affiliation{Center for Quantum Information, Institute for Interdisciplinary Information Sciences, Tsinghua University, Beijing 100084, PR China}

\author{Z.-B. Cui}
\affiliation{Center for Quantum Information, Institute for Interdisciplinary Information Sciences, Tsinghua University, Beijing 100084, PR China}

\author{R. Yao}
\affiliation{HYQ Co., Ltd., Beijing 100176, P. R. China}

\author{W.-Q. Lian}
\affiliation{HYQ Co., Ltd., Beijing 100176, P. R. China}

\author{J.-Y. Ma}
\affiliation{HYQ Co., Ltd., Beijing 100176, P. R. China}

\author{W.-X. Guo}
\affiliation{HYQ Co., Ltd., Beijing 100176, P. R. China}

\author{B.-X. Qi}
\affiliation{Center for Quantum Information, Institute for Interdisciplinary Information Sciences, Tsinghua University, Beijing 100084, PR China}

\author{P.-Y. Hou}
\affiliation{Center for Quantum Information, Institute for Interdisciplinary Information Sciences, Tsinghua University, Beijing 100084, PR China}
\affiliation{Hefei National Laboratory, Hefei 230088, PR China}

\author{Y.-F. Pu}
\affiliation{Center for Quantum Information, Institute for Interdisciplinary Information Sciences, Tsinghua University, Beijing 100084, PR China}
\affiliation{Hefei National Laboratory, Hefei 230088, PR China}

\author{Z.-C. Zhou}
\affiliation{Center for Quantum Information, Institute for Interdisciplinary Information Sciences, Tsinghua University, Beijing 100084, PR China}
\affiliation{Hefei National Laboratory, Hefei 230088, PR China}

\author{L. He}
\affiliation{Center for Quantum Information, Institute for Interdisciplinary Information Sciences, Tsinghua University, Beijing 100084, PR China}
\affiliation{Hefei National Laboratory, Hefei 230088, PR China}

\author{L.-M. Duan}
\email{Contact author: lmduan@tsinghua.edu.cn}
\affiliation{Center for Quantum Information, Institute for Interdisciplinary Information Sciences, Tsinghua University, Beijing 100084, PR China}
\affiliation{Hefei National Laboratory, Hefei 230088, PR China}

\begin{abstract}
A quantum memory is an essential element for quantum computation, quantum network and quantum metrology. Previously, a single-qubit quantum memory with a coherence time of about an hour has been realized in a dual-species setup where a coolant ion provides sympathetic cooling for a memory ion of different species. However, the frequent random position hopping between the ions in the room-temperature trap limits the technique there only applicable to single-qubit storage. Here we report a multi-ion quantum memory in a cryogenic trap based on the dual-type scheme, and demonstrate a coherence time above two hours for a logical qubit encoded in the decoherence-free subspace, i.e. two-ion entangled states, after correcting the dominant leakage error. Our scheme alleviates the necessity of an ultra-stable frequency reference for the stored qubit, and has a preferable scalability owing to the same mass of the metastable-state memory ions and the ground-state coolant ion. 
\end{abstract}

\maketitle

As the fundamental carrier of quantum information, qubits possess unique properties like superposition and entanglement, enabling wide applications beyond a classical world \cite{nielsen2000quantum}.
A quantum memory is a critical component in quantum information processing to maintain the quantum properties, namely the coherence, of the qubits \cite{lvovsky2009optical,Heshami12112016}: for quantum computing, it is necessary that the coherence time of the qubits be much longer than the execution time of elementary quantum operations for an error rate below the fault-tolerance threshold \cite{Gottesman1998,DiVincenzo2000,Ladd2010,campbell2017roads}; for quantum communication, long-lifetime quantum memories are crucial in quantum repeater protocols to efficiently establish remote quantum entanglement \cite{PhysRevLett.81.5932,duan2001long,RevModPhys.83.33}; the coherence time also limits the interrogation time of a quantum sensor and therefore its precision \cite{RevModPhys.89.035002}.

Because of its importance, various physical systems such as atomic ensembles \cite{RevModPhys.75.457,Zhao2009,PhysRevLett.114.050502,yang2016efficient,wang2019efficient,PhysRevX.14.021018}, rare-earth-doped crystals \cite{PhysRevLett.108.190503,PhysRevLett.108.190504,PhysRevLett.108.190505,Lago-Rivera2021,Ortu2022}, NV centers in diamonds \cite{doi:10.1126/science.1220513,Bar-Gill2013,PhysRevX.9.031045} and trapped ions \cite{PhysRevLett.95.060502,Haffner2005,PhysRevLett.113.220501,wang2017single,wang2021single,PhysRevA.106.062617} have been explored to realize a high-performance quantum memory. In particular, the longest coherence time about one hour has been achieved by encoding a single qubit in the hyperfine ground states of a ${}^{171}\mathrm{Yb}^+$ ion \cite{wang2017single,wang2021single}. Critical to this achievement is an ultrastable microwave frequency reference to manipulate the qubit for a long sequence of dynamical decoupling (DD) pulses. Besides, to prevent the motional heating of the ion from degrading its detection fidelity, another ion species like ${}^{137}\mathrm{Ba}^+$ is used as a coolant ion to provide sympathetic cooling for the memory ion without causing crosstalk errors. Despite this record-high storage lifetime of a single qubit, these required techniques make it challenging to generalize to a larger storage capacity for several reasons: firstly, in the room-temperature trap in Refs. \cite{wang2017single,wang2021single}, the ions frequently exchange positions with each other due to the random hopping caused by the collisions with the background gas molecules, making it impossible to retrieve information from more than one memory ions as they have random positions induced by hopping (the typical hopping time interval is much shorter than the reported coherence time in \cite{wang2017single,wang2021single}). Secondly, the global microwave can be inhomogeneous over a large ion crystal, lowering the fidelity of the DD pulses \cite{PhysRevA.106.062617}. Furthermore, to manipulate individual qubits by the global microwave, other techniques like an AC Stark shift induced by a focused laser beam \cite{PhysRevA.94.042308,doi:10.1126/sciadv.adr9527} or a Zeeman shift induced by a magnetic field gradient on magnetic-field-sensitive levels \cite{PhysRevLett.87.257904,doi:10.1126/sciadv.1600093} will be needed, leading to a decrease in the qubit coherence time. Finally, the mass difference between the memory ions and the coolant ions can also reduce the sympathetic cooling efficiency as the ion number increases \cite{PhysRevA.103.012610}.

\begin{figure*}[!tp]
   \includegraphics[width=\linewidth]{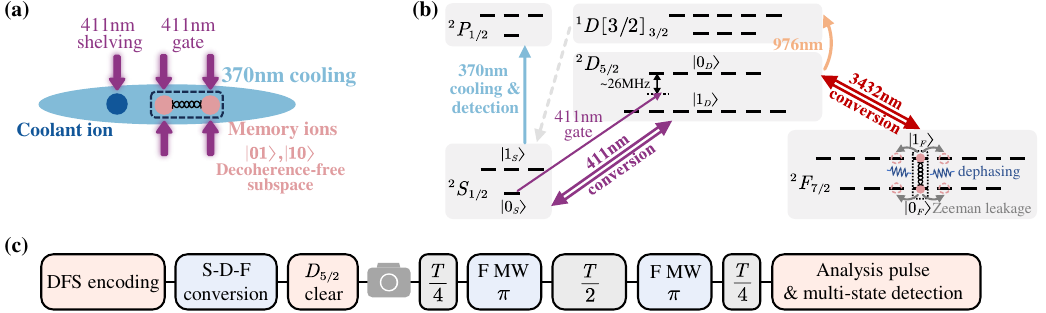}
   \caption{Experimental Scheme. (a) Three ${}^{171}\mathrm{Yb}^+$ are trapped in a linear configuration. Two neighboring ions act as memory ions whose $|0_F 1_F\rangle$ and $|1_F 0_F\rangle$ states span the decoherence-free subspace (DFS). The other ion is shined by the global $370\,$nm laser to provide sympathetic cooling. The ions can be individually addressed by $411\,$nm laser beams for tasks like entangling gates and electron shelving. (b) Relevant energy levels. The memory ions are initialized in the $S$-type for entangling gates based on focused $411\,$nm laser. Then they are converted to the $F$-type for storage via the intermediate $D_{5/2}$ levels using global bichromatic $411\,$nm and $3432\,$nm laser beams. In addition, we use global $370\,$nm laser for sympathetic cooling and for state detection, and we use global $976\,$nm laser to repump the residual population of the $D_{5/2}$ levels back to the $S_{1/2}$ levels. When stored in the DFS, the logical qubit is immune to the global dephasing error. However, it is still subjected to leakage errors to nearby Zeeman levels due to the collision with background gas molecules. (c) Experimental sequence for the memory ions. After encoding the DFS logical qubit and converting into the $F$-type, we clear the population in $D_{5/2}$ and collect the fluorescence of the ions under $370\,$nm laser (the camera symbol) to verify the successful state preparation. Then we store the DFS logical qubit in the $F$-type for time $T$, with two microwave (MW) spin echoes to further suppress the nonzero magnetic field gradient. Finally we measure the fidelity of the encoded state and the leakage error. 
   \label{fig1}}
\end{figure*}

Here we use a cryogenic ion trap \cite{Pagano_2019} to suppress the random hopping of ion positions for a multi-ion crystal and the dual-type qubit scheme \cite{yang2022realizing,10.1063/5.0069544} to solve the mass mismatch problem where memory ions and coolant ions can be mapped to the different hyperfine levels of the same ion species. Because of the level structures, in the dual-type encoding usually the coolant ions are mapped to the ground state hyperfine levels, so that the memory ions have to be mapped to other metastable levels. Previously, on-demand coherent qubit type conversions have been demonstrated and a crosstalk error rate below the fault-tolerant threshold has been reported between different qubit types \cite{yang2022realizing,Feng2024,PhysRevLett.134.070801,PhysRevLett.132.263201}. However, in terms of the storage lifetime, the current record for the metastable qubits is only about $136\,$s for the $D_{5/2}$ levels of ${}^{137}\mathrm{Ba}^+$ after discarding the leakage error \cite{PhysRevA.111.L020601}. In principle, the $F_{7/2}$ levels of ${}^{171}\mathrm{Yb}^+$ can have a lifetime above years \cite{PhysRevLett.127.213001}, but the experimentally achieved coherence time has been limited to a few seconds \cite{Feng2024}. In this work, we extend the coherence time by encoding into a decoherence-free subspace (DFS) which is robust against the dominant error source of collective spin dephasing \cite{PhysRevLett.79.1953,PhysRevLett.81.2594,doi:10.1126/science.1057357,PhysRevLett.92.220402,PhysRevLett.95.060502,Haffner2005}. This also eases the previous requirements of an ultrastable microwave frequency reference and a long DD sequence, since the frequency shift of the driving signal and the long-wavelength electromagnetic noise both act as global phases on the DFS without affecting the encoded logical qubit. Combining the dual-type qubit and the DFS encoding, we thus achieve a single-qubit coherence time above two hours after discarding the events where leakage to other $F_{7/2}$ Zeeman levels occurs. We also examine the mechanism of the leakage error and show evidence that it originates from the collision of the ions with the background gas molecules \cite{Buchachenko_2009,PhysRevLett.110.160402,PhysRevLett.117.143201}, thus pointing the way to its future suppression.

Our experimental setup is sketched in Fig.~\ref{fig1}(a) where three ${}^{171}\mathrm{Yb}^+$ ions are held in a cryogenic blade trap at a temperature of $6\,$K. One ion on the edge is used as the coolant ion under a global $370\,$nm laser, and the other two ions are memory ions to encode the DFS logical qubit. We use acousto-optic deflectors to guide focused $411\,$nm laser beams with a beam waist radius of about $2\,\upmu$m to individual ions at the separation of about $6.5\,\upmu$m. After global Doppler cooling and sideband cooling, we initialize all the three ions in $|0_S\rangle \equiv |S_{1/2},F=0,m_F=0\rangle$ by optical pumping. Then we use the focused $411\,$nm beams together with global $12.6\,$GHz microwave pulses to encode the memory ions into different DFS logical states such as $|0_S 1_S\rangle$, $|1_S 0_S\rangle$ and $(|0_S 1_S\rangle\pm |1_S 0_S\rangle)/\sqrt{2}$, where $|1_S\rangle \equiv |S_{1/2},F=1,m_F=0\rangle$. Specifically, we achieve single-qubit $R_z$ rotations by the AC Stark shift under the $411\,$nm laser, which, when sandwiched by two global microwave $\pi/2$ pulses, can prepare arbitrary computational basis states for the $S$-type qubits. Also we use counter-propagating $411\,$nm laser beams for two-qubit light-shift entangling gates \cite{PhysRevA.103.012603}. More details can be found in Supplemental Material \cite{supp}.

As shown by the energy levels in Fig.~\ref{fig1}(b), once encoded into the DFS, the two memory qubits can be converted into the $F$-type spanned by $|0_F\rangle \equiv |F_{7/2},F=3,m_F=0\rangle$ and $|1_F\rangle \equiv |F_{7/2},F=4,m_F=0\rangle$ using global bichromatic $411\,$nm and $3432\,$nm laser $\pi$ pulses. Before the conversion, we individually shelve the coolant ion into $|D_{5/2},F=2,m_F=-1\rangle$ to protect it from being affected by global conversion pulses. We add a verification step [indicated by the camera symbol in the experimental sequence in Fig.~\ref{fig1}(c)] to check if the memory ions are successfully prepared on $F$ states and the coolant ion remains on $S$ states. More details can be found in Supplemental Material \cite{supp}.

Although the DFS encoding is immune to global dephasing, a small difference in the magnetic fields felt by the two ions can still lead to a slow accumulation in the relative phase between $|0_F 1_F\rangle$ and $|1_F 0_F\rangle$ (see Supplemental Material \cite{supp}). To further suppress the effect of this magnetic field gradient and to recover the original encoded state, as shown in Fig.~\ref{fig1}(d), we apply two spin echoes by microwave $\pi$ pulses on both memory ions in the $F$-type during the total storage time of $T$. Note that this is in stark contrast to the previous works where thousands of DD pulses are needed, increasing with the storage time \cite{wang2017single,wang2021single}.

\begin{figure}[!tbp]
   \includegraphics[width=\linewidth]{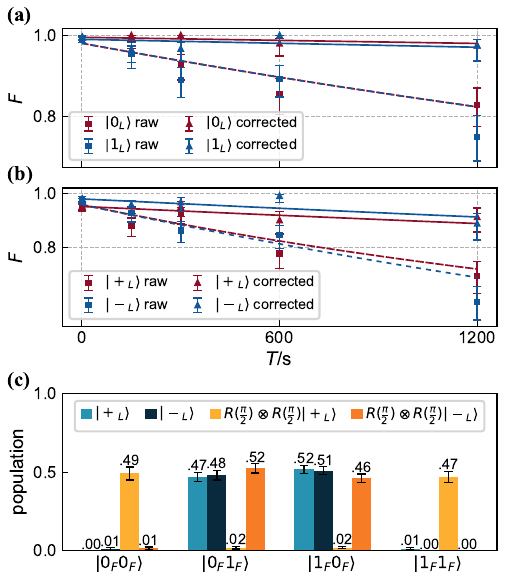}
   \caption{Storage fidelity $F$ versus storage time $T$ for the DFS logical states. In (a) and (b), the red/blue data points represent the DFS logical states $|0_L\rangle$/$|1_L\rangle$ and $|+_L\rangle$/$|-_L\rangle$, respectively. The squares denote the raw state fidelity, and the triangles are the post-selected data with leakage error detection. Error bars show $68\%$ Wilson confidence interval. The dashed curves are numerical simulation results for the leakage and the dephasing errors, and the solid curves are exponential fitting for the leakage-discarded data. (c) Two-qubit measurement probability distributions for the stored DFS logical states $|+_L\rangle$ and $|-_L\rangle$ after a storage time $T=1\,$s, and for the states with additional $\pi/2$ rotations applied on both qubits. From these populations the state fidelities can be calculated. We have normalized the distributions after discarding the leakage events.
   \label{fig2}}
\end{figure}

Finally, after a controllable storage time $T$, we measure the fidelity of the stored DFS logical state directly without decoding. Crucial to our extended storage lifetime is the capability of distinguishing the leakage error, thus leading to a heralded successful storage with a higher fidelity. To identify the possible leakage states, we perform a multi-state detection to distinguish $|0_F\rangle$ and $|1_F\rangle$ from other $F_{7/2}$ Zeeman levels (see Supplemental Material \cite{supp}). The key step of the multi-state detection is the high-fidelity population transfer from $|0_F\rangle$ to S manifolds using bichromatic $3432\,$nm $\pi$ pulses, monochromatic $411\,$nm $\pi$ pulses and $370\,$nm optical pumping pulses. Overall, we obtain a detection fidelity above $99\%$ for the $|0_F\rangle$ state and the leaked $F_{7/2}$ Zeeman levels. The fidelity for the $|1_F\rangle$ is lower at $93.4\%$ mainly due to the imperfect $3432\,$nm $\pi$ pulse. Nevertheless, such a fixed state-preparation-and-measurement (SPAM) error does not affect the measurement of the storage lifetime.

\begin{figure}[!tbp]
   \includegraphics[width=\linewidth]{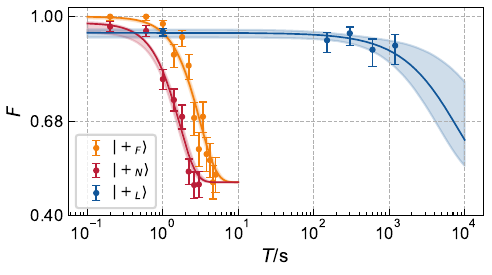}
   \caption{Storage fidelity $F$ versus storage time $T$ for different encoding schemes. The orange dots are for a single physical qubit in $|+_F\rangle$, the blue dots for the DFS logical state $|+_L\rangle\equiv(|0_F 1_F\rangle+|1_F 0_F\rangle)/\sqrt{2}$, and the red dots for the non-DFS encoding $|+_N\rangle\equiv(|0_F 0_F\rangle+|1_F 1_F\rangle)/\sqrt{2}$ which is even more sensitive to the global dephasing error than the physical qubit.
   \label{fig3}}
\end{figure}
To calibrate the storage lifetime in the DFS, we prepare four representative DFS logical states $|0_L\rangle\equiv|0_F 1_F\rangle$, $|1_L\rangle\equiv|1_F 0_F\rangle$ and $|\pm_L\rangle\equiv(|0_F 1_F\rangle\pm |1_F 0_F\rangle)/\sqrt{2}$, and measure their fidelity versus different storage time $T$ in Figs.~\ref{fig2}(a) and (b). As we can see, without post-selection, the raw fidelities of these four states (red and blue squares) decay exponentially with a $1/e$ time constant of about $2000\,$s, which is dominated by the leakage error to the nearby Zeeman levels. These data points also agree well with the numerical simulation results (dashed curves) based on a simplified error model for the leakage and the dephasing errors (see Supplemental Material \cite{supp}). On the other hand, once we discard the leakage events, the corrected state fidelities (red and blue triangles) increase and we can use the maximum likelihood method to fit an exponential decay $F=(1+A e^{-T/\tau})/2$. We obtain $\tau_0 = 3.3_{-1.7}^{+30}\times 10^4\,$s, $\tau_1 = 2.9_{-1.6}^{+46}\times 10^4\,$s, $\tau_+ = 7.9_{-3.4}^{+18.0}\times 10^3\,$s and $\tau_- = 8.0_{-3.1}^{+9.7}\times 10^3\,$s for the four DFS logical states, respectively, where the error bars represent $68\%$ confidence intervals based on Monte Carlo sampling (see Supplemental Material \cite{supp}). Note that $\tau_0$ and $\tau_1$ correspond to the storage of two-qubit product states in the computational basis and resemble the $T_1$ time for the logic qubit, while $\tau_+$ and $\tau_-$ represent the storage of two-qubit entangled states and correspond to the $T_2$ coherence time for the logic qubit.

While the fidelities of $|0_L\rangle$ and $|1_L\rangle$ can directly be measured in the computational basis, those of $|+_L\rangle$ and $|-_L\rangle$ require additional information from the superposition bases \cite{Lin2013,PhysRevLett.128.080502}. Specifically, we have $F_\pm = (\rho_{01,01}+\rho_{10,10})/2\pm\mathrm{Re}[\rho_{01,10}]$. The first term can be obtained from the population in the computational basis. As for the second term, we apply a microwave $\pi/2$ pulse on both $F$-type memory qubits with a random global phase $\phi$ which distributes uniformly between $0$ and $2\pi$. Then the probability distribution on the computational basis gives $P_{00,\pi/2}+P_{11,\pi/2}=1/2+\mathrm{Re}[\rho_{01,10}]$ and $P_{01,\pi/2}+P_{10,\pi/2}=1/2-\mathrm{Re}[\rho_{01,10}]$. We plot the measured distributions for the $|+_L\rangle$ and $|-_L\rangle$ states and those with additional $\pi/2$ pulses in Fig.~\ref{fig2}(c) at a specific storage time $T=1\,$s (normalized after discarding the leakage events). From these results, the state fidelity in Fig.~\ref{fig2}(b) can be computed.

In Fig.~\ref{fig3} we further compare the storage performance of a DFS logical state $|+_L\rangle\equiv(|0_F 1_F\rangle+ |1_F 0_F\rangle)/\sqrt{2}$ (blue), a physical state $|+_F\rangle \equiv(|0_F\rangle+|1_F\rangle)/\sqrt{2}$ (orange) and a non-DFS logical state $|+_N\rangle\equiv(|0_F 0_F\rangle+ |1_F 1_F\rangle)/\sqrt{2}$. As we can see, the DFS logical qubit has a much longer $T_2$ coherence time than the physical qubit or the non-DFS logical qubit because of its robustness against the phase noise in the driving signal and the fluctuation in the electromagnetic fields in the environment. Also we observe that the non-DFS logical state has even a shorter coherence time (about one half) than the physical qubit due to its doubled sensitivity to the dephasing errors. This result thus unambiguously proves the advantage of encoding into the DFS.

\begin{figure}[!tbp]
   \includegraphics[width=\linewidth]{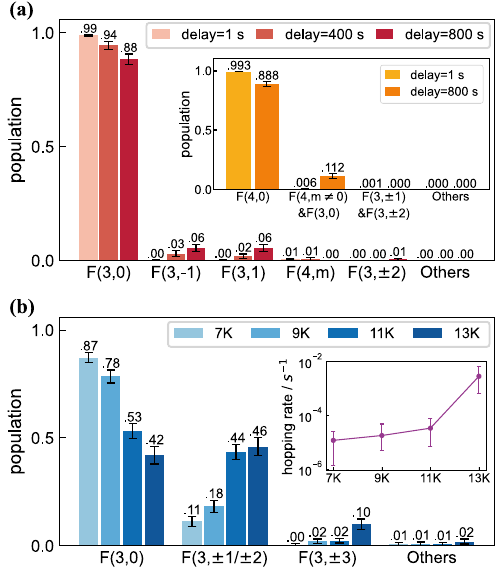}
   \caption{Analysis of leakage errors. (a) The population distribution on the $F_{7/2}$ Zeeman levels for an ion initialized in $|0_F\rangle=|F_{7/2},F=3,m_F=0\rangle$ at different storage times $T$. The leakage channel mainly goes into the nearby Zeeman levels $|F_{7/2},F=3,m_F=\pm 1\rangle$. As $T$ increases, the population of these Zeeman levels also increases. The inset shows an ion initialized in $|1_F\rangle=|F_{7/2},F=4,m_F=0\rangle$. A similar leakage probability at $T=800\,$s is observed. (b) The population distribution over the $F_{7/2}$ Zeeman levels for an ion initialized in $|0_F\rangle=|F_{7/2},F=3,m_F=0\rangle$ at different temperatures. Here we fix a storage time $T=800\,$s, and observe a larger leakage error as the temperature rises. This can be explained by the higher collision rate between the ions and the background gas molecules, which can be seen from the hopping rate of dark isotope ions in the inset.
   \label{fig4}}
\end{figure}

Finally, we try to understand the mechanism of the leakage error from the $F$-type qubit levels. Here to improve the calibration efficiency, we trap a chain of about $30$ ions and apply incomplete global conversion pulses with a rotation angle smaller than $\pi$ to probabilistically prepare about $70\%$ of the ions as the memory ions in the $F$-type, while the rest stay in the $S$-type as the coolant ions. In this way, the probability for a leakage error to occur in the whole ion chain can be amplified by about $20$ times. To determine the population distribution of the leaked states, we apply a more complicated multi-state detection sequence \cite{supp} to further distinguish different Zeeman levels of the $F_{7/2}$ manifold. In Fig.~\ref{fig4}(a) we initialize the memory ions in $|0_F\rangle=|F_{7/2},F=3,m_F=0\rangle$. As the storage time increases, the population gradually transfers into the nearby $|F_{7/2},F=3,m_F=\pm 1\rangle$ levels, while the population in other levels remains small and can be explained by the detection errors. As for the initial state $|1_F\rangle=|F_{7/2},F=4,m_F=0\rangle$, because the $|F_{7/2},F=4\rangle$, $|D_{5/2},F=3\rangle$ and $|S_{1/2},F=1\rangle$ levels have the same Zeeman splitting, it is more difficult to distinguish the population in individual Zeeman levels directly. Nevertheless, as shown in the inset of Fig.~\ref{fig4}(a), the leakage rate of $|1_F\rangle$ is similar to that of $|0_F\rangle$ after $T=800\,$s storage, suggesting the same mechanism of the leakage error.

The Zeeman leakage is sensitive to the vacuum condition of the trap, suggesting that it may originate from the collision of the ions with the background gas molecules. To further verify this hypothesis, we make use of the positive correlation between the collision rate and the temperature of the cryogenic trap \cite{Pagano_2019}. As we can see in Fig.~\ref{fig4}(b), the leakage rate increases monotonically as the temperature rises from $7\,$K to $13\,$K. At the same time, if we prepare an ion chain with a few isotope ions which appear dark under the detection laser, we observe that the hopping rate of the dark ions is also increasing monotonically with the temperature (inset), suggesting a higher collision rate. Therefore, we conclude that the leakage to the nearby Zeeman levels for the $F$-type qubits is caused by their collision with the background gas molecules, and can become a dominant error source at the storage timescale of hundreds to thousands of seconds.

In summary, we have demonstrated a coherence time above two hours for the metastable qubits by encoding into DFS logic states under the crosstalk-free sympathetic cooling of the same ion species. Such a coherence time is millions of times longer than the elementary gate operations and measurements for the dual-type qubits, and also at least tens of thousands of times longer than the typical time to generate ion-photon entanglement \cite{Feng2024}. Therefore, it paves the way toward the applications of DFS quantum memories in quantum computation and quantum network based on the dual-type encoding. Also we measure a leakage error for the $F_{7/2}$ levels to nearby Zeeman levels at a timescale of hundreds of seconds. We identify the origin of this error to be the collision of the ions with the background $\mathrm{H}_2$ gas molecules. Without correcting the leakage error, this seems to be the dominant error source for the storage fidelity. Fortunately, we demonstrate that such a leakage error can be distinguished from the desired qubit levels by a suitable multi-state detection sequence. This allows us to separate the effect of the leakage, and to treat it as an erasure error with a known location which is more friendly for quantum error correction \cite{PhysRevA.56.33,PhysRevLett.78.3217}. In the future, the Zeeman leakage can be reduced by implementing improved vacuum processing techniques.

\bigskip

\textit{Acknowledgements---}This work was supported by Innovation Program for Quantum Science and Technology (2021ZD0301601), Tsinghua University Initiative Scientific Research Program, and the Ministry of Education of China. L.M.D. acknowledges in addition support from the New Cornerstone Science Foundation through the New Cornerstone Investigator Program. Y.K.W., P.Y.H. and Y.F.P. acknowledge in addition support from the Dushi program from Tsinghua University.





\begin{thebibliography}{63}%
\makeatletter
\providecommand \@ifxundefined [1]{%
 \@ifx{#1\undefined}
}%
\providecommand \@ifnum [1]{%
 \ifnum #1\expandafter \@firstoftwo
 \else \expandafter \@secondoftwo
 \fi
}%
\providecommand \@ifx [1]{%
 \ifx #1\expandafter \@firstoftwo
 \else \expandafter \@secondoftwo
 \fi
}%
\providecommand \natexlab [1]{#1}%
\providecommand \enquote  [1]{``#1''}%
\providecommand \bibnamefont  [1]{#1}%
\providecommand \bibfnamefont [1]{#1}%
\providecommand \citenamefont [1]{#1}%
\providecommand \href@noop [0]{\@secondoftwo}%
\providecommand \href [0]{\begingroup \@sanitize@url \@href}%
\providecommand \@href[1]{\@@startlink{#1}\@@href}%
\providecommand \@@href[1]{\endgroup#1\@@endlink}%
\providecommand \@sanitize@url [0]{\catcode `\\12\catcode `\$12\catcode `\&12\catcode `\#12\catcode `\^12\catcode `\_12\catcode `\%12\relax}%
\providecommand \@@startlink[1]{}%
\providecommand \@@endlink[0]{}%
\providecommand \url  [0]{\begingroup\@sanitize@url \@url }%
\providecommand \@url [1]{\endgroup\@href {#1}{\urlprefix }}%
\providecommand \urlprefix  [0]{URL }%
\providecommand \Eprint [0]{\href }%
\providecommand \doibase [0]{http://dx.doi.org/}%
\providecommand \selectlanguage [0]{\@gobble}%
\providecommand \bibinfo  [0]{\@secondoftwo}%
\providecommand \bibfield  [0]{\@secondoftwo}%
\providecommand \translation [1]{[#1]}%
\providecommand \BibitemOpen [0]{}%
\providecommand \bibitemStop [0]{}%
\providecommand \bibitemNoStop [0]{.\EOS\space}%
\providecommand \EOS [0]{\spacefactor3000\relax}%
\providecommand \BibitemShut  [1]{\csname bibitem#1\endcsname}%
\let\auto@bib@innerbib\@empty
\bibitem [{\citenamefont {Nielsen}\ and\ \citenamefont {Chuang}(2000)}]{nielsen2000quantum}%
  \BibitemOpen
  \bibfield  {author} {\bibinfo {author} {\bibfnamefont {MA}~\bibnamefont {Nielsen}}\ and\ \bibinfo {author} {\bibfnamefont {IL}~\bibnamefont {Chuang}},\ }\href@noop {} {\emph {\bibinfo {title} {Quantum Computation and Quantum Information}}}\ (\bibinfo  {publisher} {Cambridge University Press},\ \bibinfo {year} {2000})\BibitemShut {NoStop}%
\bibitem [{\citenamefont {Lvovsky}\ \emph {et~al.}(2009)\citenamefont {Lvovsky}, \citenamefont {Sanders},\ and\ \citenamefont {Tittel}}]{lvovsky2009optical}%
  \BibitemOpen
  \bibfield  {author} {\bibinfo {author} {\bibfnamefont {Alexander~I}\ \bibnamefont {Lvovsky}}, \bibinfo {author} {\bibfnamefont {Barry~C}\ \bibnamefont {Sanders}}, \ and\ \bibinfo {author} {\bibfnamefont {Wolfgang}\ \bibnamefont {Tittel}},\ }\bibfield  {title} {\enquote {\bibinfo {title} {Optical quantum memory},}\ }\href {https://doi.org/10.1038/nphoton.2009.231} {\bibfield  {journal} {\bibinfo  {journal} {Nature photonics}\ }\textbf {\bibinfo {volume} {3}},\ \bibinfo {pages} {706--714} (\bibinfo {year} {2009})}\BibitemShut {NoStop}%
\bibitem [{\citenamefont {Heshami}\ \emph {et~al.}(2016)\citenamefont {Heshami}, \citenamefont {England}, \citenamefont {Humphreys}, \citenamefont {Bustard}, \citenamefont {Acosta}, \citenamefont {Nunn},\ and\ \citenamefont {and}}]{Heshami12112016}%
  \BibitemOpen
  \bibfield  {author} {\bibinfo {author} {\bibfnamefont {Khabat}\ \bibnamefont {Heshami}}, \bibinfo {author} {\bibfnamefont {Duncan~G.}\ \bibnamefont {England}}, \bibinfo {author} {\bibfnamefont {Peter~C.}\ \bibnamefont {Humphreys}}, \bibinfo {author} {\bibfnamefont {Philip~J.}\ \bibnamefont {Bustard}}, \bibinfo {author} {\bibfnamefont {Victor~M.}\ \bibnamefont {Acosta}}, \bibinfo {author} {\bibfnamefont {Joshua}\ \bibnamefont {Nunn}}, \ and\ \bibinfo {author} {\bibfnamefont {Benjamin J.~Sussman}\ \bibnamefont {and}},\ }\bibfield  {title} {\enquote {\bibinfo {title} {Quantum memories: emerging applications and recent advances},}\ }\href {\doibase 10.1080/09500340.2016.1148212} {\bibfield  {journal} {\bibinfo  {journal} {Journal of Modern Optics}\ }\textbf {\bibinfo {volume} {63}},\ \bibinfo {pages} {2005--2028} (\bibinfo {year} {2016})},\ \bibinfo {note} {pMID: 27695198}\BibitemShut {NoStop}%
\bibitem [{\citenamefont {Gottesman}(1998)}]{Gottesman1998}%
  \BibitemOpen
  \bibfield  {author} {\bibinfo {author} {\bibfnamefont {Daniel}\ \bibnamefont {Gottesman}},\ }\bibfield  {title} {\enquote {\bibinfo {title} {Theory of fault-tolerant quantum computation},}\ }\href {\doibase 10.1103/PhysRevA.57.127} {\bibfield  {journal} {\bibinfo  {journal} {Phys. Rev. A}\ }\textbf {\bibinfo {volume} {57}},\ \bibinfo {pages} {127--137} (\bibinfo {year} {1998})}\BibitemShut {NoStop}%
\bibitem [{\citenamefont {DiVincenzo}(2000)}]{DiVincenzo2000}%
  \BibitemOpen
  \bibfield  {author} {\bibinfo {author} {\bibfnamefont {David~P.}\ \bibnamefont {DiVincenzo}},\ }\bibfield  {title} {\enquote {\bibinfo {title} {The physical implementation of quantum computation},}\ }\href {\doibase https://doi.org/10.1002/1521-3978(200009)48:9/11<771::AID-PROP771>3.0.CO;2-E} {\bibfield  {journal} {\bibinfo  {journal} {Fortschritte der Physik}\ }\textbf {\bibinfo {volume} {48}},\ \bibinfo {pages} {771--783} (\bibinfo {year} {2000})}\BibitemShut {NoStop}%
\bibitem [{\citenamefont {Ladd}\ \emph {et~al.}(2010)\citenamefont {Ladd}, \citenamefont {Jelezko}, \citenamefont {Laflamme}, \citenamefont {Nakamura}, \citenamefont {Monroe},\ and\ \citenamefont {O'Brien}}]{Ladd2010}%
  \BibitemOpen
  \bibfield  {author} {\bibinfo {author} {\bibfnamefont {T.~D.}\ \bibnamefont {Ladd}}, \bibinfo {author} {\bibfnamefont {F.}~\bibnamefont {Jelezko}}, \bibinfo {author} {\bibfnamefont {R.}~\bibnamefont {Laflamme}}, \bibinfo {author} {\bibfnamefont {Y.}~\bibnamefont {Nakamura}}, \bibinfo {author} {\bibfnamefont {C.}~\bibnamefont {Monroe}}, \ and\ \bibinfo {author} {\bibfnamefont {J.~L.}\ \bibnamefont {O'Brien}},\ }\bibfield  {title} {\enquote {\bibinfo {title} {Quantum computers},}\ }\href {\doibase 10.1038/nature08812} {\bibfield  {journal} {\bibinfo  {journal} {Nature}\ }\textbf {\bibinfo {volume} {464}},\ \bibinfo {pages} {45--53} (\bibinfo {year} {2010})}\BibitemShut {NoStop}%
\bibitem [{\citenamefont {Campbell}\ \emph {et~al.}(2017)\citenamefont {Campbell}, \citenamefont {Terhal},\ and\ \citenamefont {Vuillot}}]{campbell2017roads}%
  \BibitemOpen
  \bibfield  {author} {\bibinfo {author} {\bibfnamefont {Earl~T}\ \bibnamefont {Campbell}}, \bibinfo {author} {\bibfnamefont {Barbara~M}\ \bibnamefont {Terhal}}, \ and\ \bibinfo {author} {\bibfnamefont {Christophe}\ \bibnamefont {Vuillot}},\ }\bibfield  {title} {\enquote {\bibinfo {title} {Roads towards fault-tolerant universal quantum computation},}\ }\href {https://doi.org/10.1038/nature23460} {\bibfield  {journal} {\bibinfo  {journal} {Nature}\ }\textbf {\bibinfo {volume} {549}},\ \bibinfo {pages} {172--179} (\bibinfo {year} {2017})}\BibitemShut {NoStop}%
\bibitem [{\citenamefont {Briegel}\ \emph {et~al.}(1998)\citenamefont {Briegel}, \citenamefont {D\"ur}, \citenamefont {Cirac},\ and\ \citenamefont {Zoller}}]{PhysRevLett.81.5932}%
  \BibitemOpen
  \bibfield  {author} {\bibinfo {author} {\bibfnamefont {H.-J.}\ \bibnamefont {Briegel}}, \bibinfo {author} {\bibfnamefont {W.}~\bibnamefont {D\"ur}}, \bibinfo {author} {\bibfnamefont {J.~I.}\ \bibnamefont {Cirac}}, \ and\ \bibinfo {author} {\bibfnamefont {P.}~\bibnamefont {Zoller}},\ }\bibfield  {title} {\enquote {\bibinfo {title} {Quantum repeaters: The role of imperfect local operations in quantum communication},}\ }\href {\doibase 10.1103/PhysRevLett.81.5932} {\bibfield  {journal} {\bibinfo  {journal} {Phys. Rev. Lett.}\ }\textbf {\bibinfo {volume} {81}},\ \bibinfo {pages} {5932--5935} (\bibinfo {year} {1998})}\BibitemShut {NoStop}%
\bibitem [{\citenamefont {Duan}\ \emph {et~al.}(2001)\citenamefont {Duan}, \citenamefont {Lukin}, \citenamefont {Cirac},\ and\ \citenamefont {Zoller}}]{duan2001long}%
  \BibitemOpen
  \bibfield  {author} {\bibinfo {author} {\bibfnamefont {L.~M.}\ \bibnamefont {Duan}}, \bibinfo {author} {\bibfnamefont {M.~D.}\ \bibnamefont {Lukin}}, \bibinfo {author} {\bibfnamefont {J.~I.}\ \bibnamefont {Cirac}}, \ and\ \bibinfo {author} {\bibfnamefont {P.}~\bibnamefont {Zoller}},\ }\bibfield  {title} {\enquote {\bibinfo {title} {Long-distance quantum communication with atomic ensembles and linear optics},}\ }\href {https://www.nature.com/articles/35106500} {\bibfield  {journal} {\bibinfo  {journal} {Nature}\ }\textbf {\bibinfo {volume} {414}},\ \bibinfo {pages} {413} (\bibinfo {year} {2001})}\BibitemShut {NoStop}%
\bibitem [{\citenamefont {Sangouard}\ \emph {et~al.}(2011)\citenamefont {Sangouard}, \citenamefont {Simon}, \citenamefont {de~Riedmatten},\ and\ \citenamefont {Gisin}}]{RevModPhys.83.33}%
  \BibitemOpen
  \bibfield  {author} {\bibinfo {author} {\bibfnamefont {Nicolas}\ \bibnamefont {Sangouard}}, \bibinfo {author} {\bibfnamefont {Christoph}\ \bibnamefont {Simon}}, \bibinfo {author} {\bibfnamefont {Hugues}\ \bibnamefont {de~Riedmatten}}, \ and\ \bibinfo {author} {\bibfnamefont {Nicolas}\ \bibnamefont {Gisin}},\ }\bibfield  {title} {\enquote {\bibinfo {title} {Quantum repeaters based on atomic ensembles and linear optics},}\ }\href {\doibase 10.1103/RevModPhys.83.33} {\bibfield  {journal} {\bibinfo  {journal} {Rev. Mod. Phys.}\ }\textbf {\bibinfo {volume} {83}},\ \bibinfo {pages} {33--80} (\bibinfo {year} {2011})}\BibitemShut {NoStop}%
\bibitem [{\citenamefont {Degen}\ \emph {et~al.}(2017)\citenamefont {Degen}, \citenamefont {Reinhard},\ and\ \citenamefont {Cappellaro}}]{RevModPhys.89.035002}%
  \BibitemOpen
  \bibfield  {author} {\bibinfo {author} {\bibfnamefont {C.~L.}\ \bibnamefont {Degen}}, \bibinfo {author} {\bibfnamefont {F.}~\bibnamefont {Reinhard}}, \ and\ \bibinfo {author} {\bibfnamefont {P.}~\bibnamefont {Cappellaro}},\ }\bibfield  {title} {\enquote {\bibinfo {title} {Quantum sensing},}\ }\href {\doibase 10.1103/RevModPhys.89.035002} {\bibfield  {journal} {\bibinfo  {journal} {Rev. Mod. Phys.}\ }\textbf {\bibinfo {volume} {89}},\ \bibinfo {pages} {035002} (\bibinfo {year} {2017})}\BibitemShut {NoStop}%
\bibitem [{\citenamefont {Lukin}(2003)}]{RevModPhys.75.457}%
  \BibitemOpen
  \bibfield  {author} {\bibinfo {author} {\bibfnamefont {M.~D.}\ \bibnamefont {Lukin}},\ }\bibfield  {title} {\enquote {\bibinfo {title} {Colloquium: Trapping and manipulating photon states in atomic ensembles},}\ }\href {\doibase 10.1103/RevModPhys.75.457} {\bibfield  {journal} {\bibinfo  {journal} {Rev. Mod. Phys.}\ }\textbf {\bibinfo {volume} {75}},\ \bibinfo {pages} {457--472} (\bibinfo {year} {2003})}\BibitemShut {NoStop}%
\bibitem [{\citenamefont {Zhao}\ \emph {et~al.}(2009)\citenamefont {Zhao}, \citenamefont {Dudin}, \citenamefont {Jenkins}, \citenamefont {Campbell}, \citenamefont {Matsukevich}, \citenamefont {Kennedy},\ and\ \citenamefont {Kuzmich}}]{Zhao2009}%
  \BibitemOpen
  \bibfield  {author} {\bibinfo {author} {\bibfnamefont {R.}~\bibnamefont {Zhao}}, \bibinfo {author} {\bibfnamefont {Y.~O.}\ \bibnamefont {Dudin}}, \bibinfo {author} {\bibfnamefont {S.~D.}\ \bibnamefont {Jenkins}}, \bibinfo {author} {\bibfnamefont {C.~J.}\ \bibnamefont {Campbell}}, \bibinfo {author} {\bibfnamefont {D.~N.}\ \bibnamefont {Matsukevich}}, \bibinfo {author} {\bibfnamefont {T.~A.~B.}\ \bibnamefont {Kennedy}}, \ and\ \bibinfo {author} {\bibfnamefont {A.}~\bibnamefont {Kuzmich}},\ }\bibfield  {title} {\enquote {\bibinfo {title} {Long-lived quantum memory},}\ }\href {\doibase 10.1038/nphys1152} {\bibfield  {journal} {\bibinfo  {journal} {Nature Physics}\ }\textbf {\bibinfo {volume} {5}},\ \bibinfo {pages} {100--104} (\bibinfo {year} {2009})}\BibitemShut {NoStop}%
\bibitem [{\citenamefont {Ding}\ \emph {et~al.}(2015)\citenamefont {Ding}, \citenamefont {Zhang}, \citenamefont {Zhou}, \citenamefont {Shi}, \citenamefont {Xiang}, \citenamefont {Wang}, \citenamefont {Jiang}, \citenamefont {Shi},\ and\ \citenamefont {Guo}}]{PhysRevLett.114.050502}%
  \BibitemOpen
  \bibfield  {author} {\bibinfo {author} {\bibfnamefont {Dong-Sheng}\ \bibnamefont {Ding}}, \bibinfo {author} {\bibfnamefont {Wei}\ \bibnamefont {Zhang}}, \bibinfo {author} {\bibfnamefont {Zhi-Yuan}\ \bibnamefont {Zhou}}, \bibinfo {author} {\bibfnamefont {Shuai}\ \bibnamefont {Shi}}, \bibinfo {author} {\bibfnamefont {Guo-Yong}\ \bibnamefont {Xiang}}, \bibinfo {author} {\bibfnamefont {Xi-Shi}\ \bibnamefont {Wang}}, \bibinfo {author} {\bibfnamefont {Yun-Kun}\ \bibnamefont {Jiang}}, \bibinfo {author} {\bibfnamefont {Bao-Sen}\ \bibnamefont {Shi}}, \ and\ \bibinfo {author} {\bibfnamefont {Guang-Can}\ \bibnamefont {Guo}},\ }\bibfield  {title} {\enquote {\bibinfo {title} {Quantum storage of orbital angular momentum entanglement in an atomic ensemble},}\ }\href {\doibase 10.1103/PhysRevLett.114.050502} {\bibfield  {journal} {\bibinfo  {journal} {Phys. Rev. Lett.}\ }\textbf {\bibinfo {volume} {114}},\ \bibinfo {pages} {050502} (\bibinfo {year} {2015})}\BibitemShut {NoStop}%
\bibitem [{\citenamefont {Yang}\ \emph {et~al.}(2016)\citenamefont {Yang}, \citenamefont {Wang}, \citenamefont {Bao},\ and\ \citenamefont {Pan}}]{yang2016efficient}%
  \BibitemOpen
  \bibfield  {author} {\bibinfo {author} {\bibfnamefont {Sheng-Jun}\ \bibnamefont {Yang}}, \bibinfo {author} {\bibfnamefont {Xu-Jie}\ \bibnamefont {Wang}}, \bibinfo {author} {\bibfnamefont {Xiao-Hui}\ \bibnamefont {Bao}}, \ and\ \bibinfo {author} {\bibfnamefont {Jian-Wei}\ \bibnamefont {Pan}},\ }\bibfield  {title} {\enquote {\bibinfo {title} {An efficient quantum light--matter interface with sub-second lifetime},}\ }\href {https://doi.org/10.1038/nphoton.2016.51} {\bibfield  {journal} {\bibinfo  {journal} {Nature Photonics}\ }\textbf {\bibinfo {volume} {10}},\ \bibinfo {pages} {381--384} (\bibinfo {year} {2016})}\BibitemShut {NoStop}%
\bibitem [{\citenamefont {Wang}\ \emph {et~al.}(2019)\citenamefont {Wang}, \citenamefont {Li}, \citenamefont {Zhang}, \citenamefont {Su}, \citenamefont {Zhou}, \citenamefont {Liao}, \citenamefont {Du}, \citenamefont {Yan},\ and\ \citenamefont {Zhu}}]{wang2019efficient}%
  \BibitemOpen
  \bibfield  {author} {\bibinfo {author} {\bibfnamefont {Yunfei}\ \bibnamefont {Wang}}, \bibinfo {author} {\bibfnamefont {Jianfeng}\ \bibnamefont {Li}}, \bibinfo {author} {\bibfnamefont {Shanchao}\ \bibnamefont {Zhang}}, \bibinfo {author} {\bibfnamefont {Keyu}\ \bibnamefont {Su}}, \bibinfo {author} {\bibfnamefont {Yiru}\ \bibnamefont {Zhou}}, \bibinfo {author} {\bibfnamefont {Kaiyu}\ \bibnamefont {Liao}}, \bibinfo {author} {\bibfnamefont {Shengwang}\ \bibnamefont {Du}}, \bibinfo {author} {\bibfnamefont {Hui}\ \bibnamefont {Yan}}, \ and\ \bibinfo {author} {\bibfnamefont {Shi-Liang}\ \bibnamefont {Zhu}},\ }\bibfield  {title} {\enquote {\bibinfo {title} {Efficient quantum memory for single-photon polarization qubits},}\ }\href {https://doi.org/10.1038/s41566-019-0368-8} {\bibfield  {journal} {\bibinfo  {journal} {Nature Photonics}\ }\textbf {\bibinfo {volume} {13}},\ \bibinfo {pages} {346--351} (\bibinfo {year} {2019})}\BibitemShut {NoStop}%
\bibitem [{\citenamefont {Zhang}\ \emph {et~al.}(2024)\citenamefont {Zhang}, \citenamefont {Shi}, \citenamefont {Cui}, \citenamefont {Wang}, \citenamefont {Wu}, \citenamefont {Duan},\ and\ \citenamefont {Pu}}]{PhysRevX.14.021018}%
  \BibitemOpen
  \bibfield  {author} {\bibinfo {author} {\bibfnamefont {Sheng}\ \bibnamefont {Zhang}}, \bibinfo {author} {\bibfnamefont {Jixuan}\ \bibnamefont {Shi}}, \bibinfo {author} {\bibfnamefont {Zhaibin}\ \bibnamefont {Cui}}, \bibinfo {author} {\bibfnamefont {Ye}~\bibnamefont {Wang}}, \bibinfo {author} {\bibfnamefont {Yukai}\ \bibnamefont {Wu}}, \bibinfo {author} {\bibfnamefont {Luming}\ \bibnamefont {Duan}}, \ and\ \bibinfo {author} {\bibfnamefont {Yunfei}\ \bibnamefont {Pu}},\ }\bibfield  {title} {\enquote {\bibinfo {title} {Realization of a programmable multipurpose photonic quantum memory with over-thousand qubit manipulations},}\ }\href {\doibase 10.1103/PhysRevX.14.021018} {\bibfield  {journal} {\bibinfo  {journal} {Phys. Rev. X}\ }\textbf {\bibinfo {volume} {14}},\ \bibinfo {pages} {021018} (\bibinfo {year} {2024})}\BibitemShut {NoStop}%
\bibitem [{\citenamefont {Clausen}\ \emph {et~al.}(2012)\citenamefont {Clausen}, \citenamefont {Bussi\`eres}, \citenamefont {Afzelius},\ and\ \citenamefont {Gisin}}]{PhysRevLett.108.190503}%
  \BibitemOpen
  \bibfield  {author} {\bibinfo {author} {\bibfnamefont {Christoph}\ \bibnamefont {Clausen}}, \bibinfo {author} {\bibfnamefont {F\'elix}\ \bibnamefont {Bussi\`eres}}, \bibinfo {author} {\bibfnamefont {Mikael}\ \bibnamefont {Afzelius}}, \ and\ \bibinfo {author} {\bibfnamefont {Nicolas}\ \bibnamefont {Gisin}},\ }\bibfield  {title} {\enquote {\bibinfo {title} {Quantum storage of heralded polarization qubits in birefringent and anisotropically absorbing materials},}\ }\href {\doibase 10.1103/PhysRevLett.108.190503} {\bibfield  {journal} {\bibinfo  {journal} {Phys. Rev. Lett.}\ }\textbf {\bibinfo {volume} {108}},\ \bibinfo {pages} {190503} (\bibinfo {year} {2012})}\BibitemShut {NoStop}%
\bibitem [{\citenamefont {G\"undo\ifmmode~\breve{g}\else \u{g}\fi{}an}\ \emph {et~al.}(2012)\citenamefont {G\"undo\ifmmode~\breve{g}\else \u{g}\fi{}an}, \citenamefont {Ledingham}, \citenamefont {Almasi}, \citenamefont {Cristiani},\ and\ \citenamefont {de~Riedmatten}}]{PhysRevLett.108.190504}%
  \BibitemOpen
  \bibfield  {author} {\bibinfo {author} {\bibfnamefont {Mustafa}\ \bibnamefont {G\"undo\ifmmode~\breve{g}\else \u{g}\fi{}an}}, \bibinfo {author} {\bibfnamefont {Patrick~M.}\ \bibnamefont {Ledingham}}, \bibinfo {author} {\bibfnamefont {Attaallah}\ \bibnamefont {Almasi}}, \bibinfo {author} {\bibfnamefont {Matteo}\ \bibnamefont {Cristiani}}, \ and\ \bibinfo {author} {\bibfnamefont {Hugues}\ \bibnamefont {de~Riedmatten}},\ }\bibfield  {title} {\enquote {\bibinfo {title} {Quantum storage of a photonic polarization qubit in a solid},}\ }\href {\doibase 10.1103/PhysRevLett.108.190504} {\bibfield  {journal} {\bibinfo  {journal} {Phys. Rev. Lett.}\ }\textbf {\bibinfo {volume} {108}},\ \bibinfo {pages} {190504} (\bibinfo {year} {2012})}\BibitemShut {NoStop}%
\bibitem [{\citenamefont {Zhou}\ \emph {et~al.}(2012)\citenamefont {Zhou}, \citenamefont {Lin}, \citenamefont {Yang}, \citenamefont {Li},\ and\ \citenamefont {Guo}}]{PhysRevLett.108.190505}%
  \BibitemOpen
  \bibfield  {author} {\bibinfo {author} {\bibfnamefont {Zong-Quan}\ \bibnamefont {Zhou}}, \bibinfo {author} {\bibfnamefont {Wei-Bin}\ \bibnamefont {Lin}}, \bibinfo {author} {\bibfnamefont {Ming}\ \bibnamefont {Yang}}, \bibinfo {author} {\bibfnamefont {Chuan-Feng}\ \bibnamefont {Li}}, \ and\ \bibinfo {author} {\bibfnamefont {Guang-Can}\ \bibnamefont {Guo}},\ }\bibfield  {title} {\enquote {\bibinfo {title} {Realization of reliable solid-state quantum memory for photonic polarization qubit},}\ }\href {\doibase 10.1103/PhysRevLett.108.190505} {\bibfield  {journal} {\bibinfo  {journal} {Phys. Rev. Lett.}\ }\textbf {\bibinfo {volume} {108}},\ \bibinfo {pages} {190505} (\bibinfo {year} {2012})}\BibitemShut {NoStop}%
\bibitem [{\citenamefont {Lago-Rivera}\ \emph {et~al.}(2021)\citenamefont {Lago-Rivera}, \citenamefont {Grandi}, \citenamefont {Rakonjac}, \citenamefont {Seri},\ and\ \citenamefont {de~Riedmatten}}]{Lago-Rivera2021}%
  \BibitemOpen
  \bibfield  {author} {\bibinfo {author} {\bibfnamefont {Dario}\ \bibnamefont {Lago-Rivera}}, \bibinfo {author} {\bibfnamefont {Samuele}\ \bibnamefont {Grandi}}, \bibinfo {author} {\bibfnamefont {Jelena~V.}\ \bibnamefont {Rakonjac}}, \bibinfo {author} {\bibfnamefont {Alessandro}\ \bibnamefont {Seri}}, \ and\ \bibinfo {author} {\bibfnamefont {Hugues}\ \bibnamefont {de~Riedmatten}},\ }\bibfield  {title} {\enquote {\bibinfo {title} {Telecom-heralded entanglement between multimode solid-state quantum memories},}\ }\href {\doibase 10.1038/s41586-021-03481-8} {\bibfield  {journal} {\bibinfo  {journal} {Nature}\ }\textbf {\bibinfo {volume} {594}},\ \bibinfo {pages} {37--40} (\bibinfo {year} {2021})}\BibitemShut {NoStop}%
\bibitem [{\citenamefont {Ortu}\ \emph {et~al.}(2022)\citenamefont {Ortu}, \citenamefont {Holz{\"a}pfel}, \citenamefont {Etesse},\ and\ \citenamefont {Afzelius}}]{Ortu2022}%
  \BibitemOpen
  \bibfield  {author} {\bibinfo {author} {\bibfnamefont {Antonio}\ \bibnamefont {Ortu}}, \bibinfo {author} {\bibfnamefont {Adrian}\ \bibnamefont {Holz{\"a}pfel}}, \bibinfo {author} {\bibfnamefont {Jean}\ \bibnamefont {Etesse}}, \ and\ \bibinfo {author} {\bibfnamefont {Mikael}\ \bibnamefont {Afzelius}},\ }\bibfield  {title} {\enquote {\bibinfo {title} {Storage of photonic time-bin qubits for up to 20{\thinspace}ms in a rare-earth doped crystal},}\ }\href {\doibase 10.1038/s41534-022-00541-3} {\bibfield  {journal} {\bibinfo  {journal} {npj Quantum Information}\ }\textbf {\bibinfo {volume} {8}},\ \bibinfo {pages} {29} (\bibinfo {year} {2022})}\BibitemShut {NoStop}%
\bibitem [{\citenamefont {Maurer}\ \emph {et~al.}(2012)\citenamefont {Maurer}, \citenamefont {Kucsko}, \citenamefont {Latta}, \citenamefont {Jiang}, \citenamefont {Yao}, \citenamefont {Bennett}, \citenamefont {Pastawski}, \citenamefont {Hunger}, \citenamefont {Chisholm}, \citenamefont {Markham}, \citenamefont {Twitchen}, \citenamefont {Cirac},\ and\ \citenamefont {Lukin}}]{doi:10.1126/science.1220513}%
  \BibitemOpen
  \bibfield  {author} {\bibinfo {author} {\bibfnamefont {P.~C.}\ \bibnamefont {Maurer}}, \bibinfo {author} {\bibfnamefont {G.}~\bibnamefont {Kucsko}}, \bibinfo {author} {\bibfnamefont {C.}~\bibnamefont {Latta}}, \bibinfo {author} {\bibfnamefont {L.}~\bibnamefont {Jiang}}, \bibinfo {author} {\bibfnamefont {N.~Y.}\ \bibnamefont {Yao}}, \bibinfo {author} {\bibfnamefont {S.~D.}\ \bibnamefont {Bennett}}, \bibinfo {author} {\bibfnamefont {F.}~\bibnamefont {Pastawski}}, \bibinfo {author} {\bibfnamefont {D.}~\bibnamefont {Hunger}}, \bibinfo {author} {\bibfnamefont {N.}~\bibnamefont {Chisholm}}, \bibinfo {author} {\bibfnamefont {M.}~\bibnamefont {Markham}}, \bibinfo {author} {\bibfnamefont {D.~J.}\ \bibnamefont {Twitchen}}, \bibinfo {author} {\bibfnamefont {J.~I.}\ \bibnamefont {Cirac}}, \ and\ \bibinfo {author} {\bibfnamefont {M.~D.}\ \bibnamefont {Lukin}},\ }\bibfield  {title} {\enquote {\bibinfo {title} {Room-temperature quantum bit memory exceeding one second},}\ }\href {\doibase 10.1126/science.1220513} {\bibfield  {journal} {\bibinfo  {journal} {Science}\ }\textbf {\bibinfo {volume} {336}},\ \bibinfo {pages} {1283--1286} (\bibinfo {year} {2012})}\BibitemShut {NoStop}%
\bibitem [{\citenamefont {Bar-Gill}\ \emph {et~al.}(2013)\citenamefont {Bar-Gill}, \citenamefont {Pham}, \citenamefont {Jarmola}, \citenamefont {Budker},\ and\ \citenamefont {Walsworth}}]{Bar-Gill2013}%
  \BibitemOpen
  \bibfield  {author} {\bibinfo {author} {\bibfnamefont {N.}~\bibnamefont {Bar-Gill}}, \bibinfo {author} {\bibfnamefont {L.~M.}\ \bibnamefont {Pham}}, \bibinfo {author} {\bibfnamefont {A.}~\bibnamefont {Jarmola}}, \bibinfo {author} {\bibfnamefont {D.}~\bibnamefont {Budker}}, \ and\ \bibinfo {author} {\bibfnamefont {R.~L.}\ \bibnamefont {Walsworth}},\ }\bibfield  {title} {\enquote {\bibinfo {title} {Solid-state electronic spin coherence time approaching one second},}\ }\href {\doibase 10.1038/ncomms2771} {\bibfield  {journal} {\bibinfo  {journal} {Nature Communications}\ }\textbf {\bibinfo {volume} {4}},\ \bibinfo {pages} {1743} (\bibinfo {year} {2013})}\BibitemShut {NoStop}%
\bibitem [{\citenamefont {Bradley}\ \emph {et~al.}(2019)\citenamefont {Bradley}, \citenamefont {Randall}, \citenamefont {Abobeih}, \citenamefont {Berrevoets}, \citenamefont {Degen}, \citenamefont {Bakker}, \citenamefont {Markham}, \citenamefont {Twitchen},\ and\ \citenamefont {Taminiau}}]{PhysRevX.9.031045}%
  \BibitemOpen
  \bibfield  {author} {\bibinfo {author} {\bibfnamefont {C.~E.}\ \bibnamefont {Bradley}}, \bibinfo {author} {\bibfnamefont {J.}~\bibnamefont {Randall}}, \bibinfo {author} {\bibfnamefont {M.~H.}\ \bibnamefont {Abobeih}}, \bibinfo {author} {\bibfnamefont {R.~C.}\ \bibnamefont {Berrevoets}}, \bibinfo {author} {\bibfnamefont {M.~J.}\ \bibnamefont {Degen}}, \bibinfo {author} {\bibfnamefont {M.~A.}\ \bibnamefont {Bakker}}, \bibinfo {author} {\bibfnamefont {M.}~\bibnamefont {Markham}}, \bibinfo {author} {\bibfnamefont {D.~J.}\ \bibnamefont {Twitchen}}, \ and\ \bibinfo {author} {\bibfnamefont {T.~H.}\ \bibnamefont {Taminiau}},\ }\bibfield  {title} {\enquote {\bibinfo {title} {A ten-qubit solid-state spin register with quantum memory up to one minute},}\ }\href {\doibase 10.1103/PhysRevX.9.031045} {\bibfield  {journal} {\bibinfo  {journal} {Phys. Rev. X}\ }\textbf {\bibinfo {volume} {9}},\ \bibinfo {pages} {031045} (\bibinfo {year} {2019})}\BibitemShut {NoStop}%
\bibitem [{\citenamefont {Langer}\ \emph {et~al.}(2005)\citenamefont {Langer}, \citenamefont {Ozeri}, \citenamefont {Jost}, \citenamefont {Chiaverini}, \citenamefont {DeMarco}, \citenamefont {Ben-Kish}, \citenamefont {Blakestad}, \citenamefont {Britton}, \citenamefont {Hume}, \citenamefont {Itano}, \citenamefont {Leibfried}, \citenamefont {Reichle}, \citenamefont {Rosenband}, \citenamefont {Schaetz}, \citenamefont {Schmidt},\ and\ \citenamefont {Wineland}}]{PhysRevLett.95.060502}%
  \BibitemOpen
  \bibfield  {author} {\bibinfo {author} {\bibfnamefont {C.}~\bibnamefont {Langer}}, \bibinfo {author} {\bibfnamefont {R.}~\bibnamefont {Ozeri}}, \bibinfo {author} {\bibfnamefont {J.~D.}\ \bibnamefont {Jost}}, \bibinfo {author} {\bibfnamefont {J.}~\bibnamefont {Chiaverini}}, \bibinfo {author} {\bibfnamefont {B.}~\bibnamefont {DeMarco}}, \bibinfo {author} {\bibfnamefont {A.}~\bibnamefont {Ben-Kish}}, \bibinfo {author} {\bibfnamefont {R.~B.}\ \bibnamefont {Blakestad}}, \bibinfo {author} {\bibfnamefont {J.}~\bibnamefont {Britton}}, \bibinfo {author} {\bibfnamefont {D.~B.}\ \bibnamefont {Hume}}, \bibinfo {author} {\bibfnamefont {W.~M.}\ \bibnamefont {Itano}}, \bibinfo {author} {\bibfnamefont {D.}~\bibnamefont {Leibfried}}, \bibinfo {author} {\bibfnamefont {R.}~\bibnamefont {Reichle}}, \bibinfo {author} {\bibfnamefont {T.}~\bibnamefont {Rosenband}}, \bibinfo {author} {\bibfnamefont {T.}~\bibnamefont {Schaetz}}, \bibinfo {author} {\bibfnamefont {P.~O.}\ \bibnamefont {Schmidt}}, \ and\ \bibinfo {author} {\bibfnamefont {D.~J.}\ \bibnamefont {Wineland}},\ }\bibfield  {title} {\enquote {\bibinfo {title} {Long-lived qubit memory using atomic ions},}\ }\href {\doibase 10.1103/PhysRevLett.95.060502} {\bibfield  {journal} {\bibinfo  {journal} {Phys. Rev. Lett.}\ }\textbf {\bibinfo {volume} {95}},\ \bibinfo {pages} {060502} (\bibinfo {year} {2005})}\BibitemShut {NoStop}%
\bibitem [{\citenamefont {H{\"a}ffner}\ \emph {et~al.}(2005)\citenamefont {H{\"a}ffner}, \citenamefont {Schmidt-Kaler}, \citenamefont {H{\"a}nsel}, \citenamefont {Roos}, \citenamefont {K{\"o}rber}, \citenamefont {Chwalla}, \citenamefont {Riebe}, \citenamefont {Benhelm}, \citenamefont {Rapol}, \citenamefont {Becher},\ and\ \citenamefont {Blatt}}]{Haffner2005}%
  \BibitemOpen
  \bibfield  {author} {\bibinfo {author} {\bibfnamefont {H.}~\bibnamefont {H{\"a}ffner}}, \bibinfo {author} {\bibfnamefont {F.}~\bibnamefont {Schmidt-Kaler}}, \bibinfo {author} {\bibfnamefont {W.}~\bibnamefont {H{\"a}nsel}}, \bibinfo {author} {\bibfnamefont {C.~F.}\ \bibnamefont {Roos}}, \bibinfo {author} {\bibfnamefont {T.}~\bibnamefont {K{\"o}rber}}, \bibinfo {author} {\bibfnamefont {M.}~\bibnamefont {Chwalla}}, \bibinfo {author} {\bibfnamefont {M.}~\bibnamefont {Riebe}}, \bibinfo {author} {\bibfnamefont {J.}~\bibnamefont {Benhelm}}, \bibinfo {author} {\bibfnamefont {U.~D.}\ \bibnamefont {Rapol}}, \bibinfo {author} {\bibfnamefont {C.}~\bibnamefont {Becher}}, \ and\ \bibinfo {author} {\bibfnamefont {R.}~\bibnamefont {Blatt}},\ }\bibfield  {title} {\enquote {\bibinfo {title} {Robust entanglement},}\ }\href {\doibase 10.1007/s00340-005-1917-z} {\bibfield  {journal} {\bibinfo  {journal} {Applied Physics B}\ }\textbf {\bibinfo {volume} {81}},\ \bibinfo {pages} {151--153} (\bibinfo {year} {2005})}\BibitemShut {NoStop}%
\bibitem [{\citenamefont {Harty}\ \emph {et~al.}(2014)\citenamefont {Harty}, \citenamefont {Allcock}, \citenamefont {Ballance}, \citenamefont {Guidoni}, \citenamefont {Janacek}, \citenamefont {Linke}, \citenamefont {Stacey},\ and\ \citenamefont {Lucas}}]{PhysRevLett.113.220501}%
  \BibitemOpen
  \bibfield  {author} {\bibinfo {author} {\bibfnamefont {T.~P.}\ \bibnamefont {Harty}}, \bibinfo {author} {\bibfnamefont {D.~T.~C.}\ \bibnamefont {Allcock}}, \bibinfo {author} {\bibfnamefont {C.~J.}\ \bibnamefont {Ballance}}, \bibinfo {author} {\bibfnamefont {L.}~\bibnamefont {Guidoni}}, \bibinfo {author} {\bibfnamefont {H.~A.}\ \bibnamefont {Janacek}}, \bibinfo {author} {\bibfnamefont {N.~M.}\ \bibnamefont {Linke}}, \bibinfo {author} {\bibfnamefont {D.~N.}\ \bibnamefont {Stacey}}, \ and\ \bibinfo {author} {\bibfnamefont {D.~M.}\ \bibnamefont {Lucas}},\ }\bibfield  {title} {\enquote {\bibinfo {title} {High-fidelity preparation, gates, memory, and readout of a trapped-ion quantum bit},}\ }\href {\doibase 10.1103/PhysRevLett.113.220501} {\bibfield  {journal} {\bibinfo  {journal} {Phys. Rev. Lett.}\ }\textbf {\bibinfo {volume} {113}},\ \bibinfo {pages} {220501} (\bibinfo {year} {2014})}\BibitemShut {NoStop}%
\bibitem [{\citenamefont {Wang}\ \emph {et~al.}(2017)\citenamefont {Wang}, \citenamefont {Um}, \citenamefont {Zhang}, \citenamefont {An}, \citenamefont {Lyu}, \citenamefont {Zhang}, \citenamefont {Duan}, \citenamefont {Yum},\ and\ \citenamefont {Kim}}]{wang2017single}%
  \BibitemOpen
  \bibfield  {author} {\bibinfo {author} {\bibfnamefont {Ye}~\bibnamefont {Wang}}, \bibinfo {author} {\bibfnamefont {Mark}\ \bibnamefont {Um}}, \bibinfo {author} {\bibfnamefont {Junhua}\ \bibnamefont {Zhang}}, \bibinfo {author} {\bibfnamefont {Shuoming}\ \bibnamefont {An}}, \bibinfo {author} {\bibfnamefont {Ming}\ \bibnamefont {Lyu}}, \bibinfo {author} {\bibfnamefont {Jing-Ning}\ \bibnamefont {Zhang}}, \bibinfo {author} {\bibfnamefont {L.-M.}\ \bibnamefont {Duan}}, \bibinfo {author} {\bibfnamefont {Dahyun}\ \bibnamefont {Yum}}, \ and\ \bibinfo {author} {\bibfnamefont {Kihwan}\ \bibnamefont {Kim}},\ }\bibfield  {title} {\enquote {\bibinfo {title} {Single-qubit quantum memory exceeding ten-minute coherence time},}\ }\href {https://doi.org/10.1038/s41566-017-0007-1} {\bibfield  {journal} {\bibinfo  {journal} {Nature photonics}\ }\textbf {\bibinfo {volume} {11}},\ \bibinfo {pages} {646--650} (\bibinfo {year} {2017})}\BibitemShut {NoStop}%
\bibitem [{\citenamefont {Wang}\ \emph {et~al.}(2021)\citenamefont {Wang}, \citenamefont {Luan}, \citenamefont {Qiao}, \citenamefont {Um}, \citenamefont {Zhang}, \citenamefont {Wang}, \citenamefont {Yuan}, \citenamefont {Gu}, \citenamefont {Zhang},\ and\ \citenamefont {Kim}}]{wang2021single}%
  \BibitemOpen
  \bibfield  {author} {\bibinfo {author} {\bibfnamefont {Pengfei}\ \bibnamefont {Wang}}, \bibinfo {author} {\bibfnamefont {Chun-Yang}\ \bibnamefont {Luan}}, \bibinfo {author} {\bibfnamefont {Mu}~\bibnamefont {Qiao}}, \bibinfo {author} {\bibfnamefont {Mark}\ \bibnamefont {Um}}, \bibinfo {author} {\bibfnamefont {Junhua}\ \bibnamefont {Zhang}}, \bibinfo {author} {\bibfnamefont {Ye}~\bibnamefont {Wang}}, \bibinfo {author} {\bibfnamefont {Xiao}\ \bibnamefont {Yuan}}, \bibinfo {author} {\bibfnamefont {Mile}\ \bibnamefont {Gu}}, \bibinfo {author} {\bibfnamefont {Jingning}\ \bibnamefont {Zhang}}, \ and\ \bibinfo {author} {\bibfnamefont {Kihwan}\ \bibnamefont {Kim}},\ }\bibfield  {title} {\enquote {\bibinfo {title} {Single ion qubit with estimated coherence time exceeding one hour},}\ }\href {https://doi.org/10.1038/s41467-020-20330-w} {\bibfield  {journal} {\bibinfo  {journal} {Nature communications}\ }\textbf {\bibinfo {volume} {12}},\ \bibinfo {pages} {233} (\bibinfo {year} {2021})}\BibitemShut {NoStop}%
\bibitem [{\citenamefont {Yao}\ \emph {et~al.}(2022)\citenamefont {Yao}, \citenamefont {Lian}, \citenamefont {Wu}, \citenamefont {Wang}, \citenamefont {Li}, \citenamefont {Mei}, \citenamefont {Qi}, \citenamefont {Yao}, \citenamefont {Zhou}, \citenamefont {He},\ and\ \citenamefont {Duan}}]{PhysRevA.106.062617}%
  \BibitemOpen
  \bibfield  {author} {\bibinfo {author} {\bibfnamefont {R.}~\bibnamefont {Yao}}, \bibinfo {author} {\bibfnamefont {W.-Q.}\ \bibnamefont {Lian}}, \bibinfo {author} {\bibfnamefont {Y.-K.}\ \bibnamefont {Wu}}, \bibinfo {author} {\bibfnamefont {G.-X.}\ \bibnamefont {Wang}}, \bibinfo {author} {\bibfnamefont {B.-W.}\ \bibnamefont {Li}}, \bibinfo {author} {\bibfnamefont {Q.-X.}\ \bibnamefont {Mei}}, \bibinfo {author} {\bibfnamefont {B.-X.}\ \bibnamefont {Qi}}, \bibinfo {author} {\bibfnamefont {L.}~\bibnamefont {Yao}}, \bibinfo {author} {\bibfnamefont {Z.-C.}\ \bibnamefont {Zhou}}, \bibinfo {author} {\bibfnamefont {L.}~\bibnamefont {He}}, \ and\ \bibinfo {author} {\bibfnamefont {L.-M.}\ \bibnamefont {Duan}},\ }\bibfield  {title} {\enquote {\bibinfo {title} {Experimental realization of a multiqubit quantum memory in a 218-ion chain},}\ }\href {\doibase 10.1103/PhysRevA.106.062617} {\bibfield  {journal} {\bibinfo  {journal} {Phys. Rev. A}\ }\textbf {\bibinfo {volume} {106}},\ \bibinfo {pages} {062617} (\bibinfo {year} {2022})}\BibitemShut {NoStop}%
\bibitem [{\citenamefont {Lee}\ \emph {et~al.}(2016)\citenamefont {Lee}, \citenamefont {Smith}, \citenamefont {Richerme}, \citenamefont {Neyenhuis}, \citenamefont {Hess}, \citenamefont {Zhang},\ and\ \citenamefont {Monroe}}]{PhysRevA.94.042308}%
  \BibitemOpen
  \bibfield  {author} {\bibinfo {author} {\bibfnamefont {A.~C.}\ \bibnamefont {Lee}}, \bibinfo {author} {\bibfnamefont {J.}~\bibnamefont {Smith}}, \bibinfo {author} {\bibfnamefont {P.}~\bibnamefont {Richerme}}, \bibinfo {author} {\bibfnamefont {B.}~\bibnamefont {Neyenhuis}}, \bibinfo {author} {\bibfnamefont {P.~W.}\ \bibnamefont {Hess}}, \bibinfo {author} {\bibfnamefont {J.}~\bibnamefont {Zhang}}, \ and\ \bibinfo {author} {\bibfnamefont {C.}~\bibnamefont {Monroe}},\ }\bibfield  {title} {\enquote {\bibinfo {title} {Engineering large stark shifts for control of individual clock state qubits},}\ }\href {\doibase 10.1103/PhysRevA.94.042308} {\bibfield  {journal} {\bibinfo  {journal} {Phys. Rev. A}\ }\textbf {\bibinfo {volume} {94}},\ \bibinfo {pages} {042308} (\bibinfo {year} {2016})}\BibitemShut {NoStop}%
\bibitem [{\citenamefont {Cheng}\ \emph {et~al.}(2024)\citenamefont {Cheng}, \citenamefont {Wu}, \citenamefont {Li}, \citenamefont {Mei}, \citenamefont {Li}, \citenamefont {Wang}, \citenamefont {Jiang}, \citenamefont {Qi}, \citenamefont {Zhou}, \citenamefont {Hou},\ and\ \citenamefont {Duan}}]{doi:10.1126/sciadv.adr9527}%
  \BibitemOpen
  \bibfield  {author} {\bibinfo {author} {\bibfnamefont {Zhi-Jie}\ \bibnamefont {Cheng}}, \bibinfo {author} {\bibfnamefont {Yu-Kai}\ \bibnamefont {Wu}}, \bibinfo {author} {\bibfnamefont {Shi-Jiao}\ \bibnamefont {Li}}, \bibinfo {author} {\bibfnamefont {Quan-Xin}\ \bibnamefont {Mei}}, \bibinfo {author} {\bibfnamefont {Bo-Wen}\ \bibnamefont {Li}}, \bibinfo {author} {\bibfnamefont {Gang-Xi}\ \bibnamefont {Wang}}, \bibinfo {author} {\bibfnamefont {Yue}\ \bibnamefont {Jiang}}, \bibinfo {author} {\bibfnamefont {Bin-Xiang}\ \bibnamefont {Qi}}, \bibinfo {author} {\bibfnamefont {Zi-Chao}\ \bibnamefont {Zhou}}, \bibinfo {author} {\bibfnamefont {Pan-Yu}\ \bibnamefont {Hou}}, \ and\ \bibinfo {author} {\bibfnamefont {Lu-Ming}\ \bibnamefont {Duan}},\ }\bibfield  {title} {\enquote {\bibinfo {title} {Observation of quantum superposition of topological defects in a trapped-ion quantum simulator},}\ }\href {\doibase 10.1126/sciadv.adr9527} {\bibfield  {journal} {\bibinfo  {journal} {Science Advances}\ }\textbf {\bibinfo {volume} {10}},\ \bibinfo {pages} {eadr9527} (\bibinfo {year} {2024})}\BibitemShut {NoStop}%
\bibitem [{\citenamefont {Mintert}\ and\ \citenamefont {Wunderlich}(2001)}]{PhysRevLett.87.257904}%
  \BibitemOpen
  \bibfield  {author} {\bibinfo {author} {\bibfnamefont {Florian}\ \bibnamefont {Mintert}}\ and\ \bibinfo {author} {\bibfnamefont {Christof}\ \bibnamefont {Wunderlich}},\ }\bibfield  {title} {\enquote {\bibinfo {title} {Ion-trap quantum logic using long-wavelength radiation},}\ }\href {\doibase 10.1103/PhysRevLett.87.257904} {\bibfield  {journal} {\bibinfo  {journal} {Phys. Rev. Lett.}\ }\textbf {\bibinfo {volume} {87}},\ \bibinfo {pages} {257904} (\bibinfo {year} {2001})}\BibitemShut {NoStop}%
\bibitem [{\citenamefont {Piltz}\ \emph {et~al.}(2016)\citenamefont {Piltz}, \citenamefont {Sriarunothai}, \citenamefont {Ivanov}, \citenamefont {Wölk},\ and\ \citenamefont {Wunderlich}}]{doi:10.1126/sciadv.1600093}%
  \BibitemOpen
  \bibfield  {author} {\bibinfo {author} {\bibfnamefont {Christian}\ \bibnamefont {Piltz}}, \bibinfo {author} {\bibfnamefont {Theeraphot}\ \bibnamefont {Sriarunothai}}, \bibinfo {author} {\bibfnamefont {Svetoslav~S.}\ \bibnamefont {Ivanov}}, \bibinfo {author} {\bibfnamefont {Sabine}\ \bibnamefont {Wölk}}, \ and\ \bibinfo {author} {\bibfnamefont {Christof}\ \bibnamefont {Wunderlich}},\ }\bibfield  {title} {\enquote {\bibinfo {title} {Versatile microwave-driven trapped ion spin system for quantum information processing},}\ }\href {\doibase 10.1126/sciadv.1600093} {\bibfield  {journal} {\bibinfo  {journal} {Science Advances}\ }\textbf {\bibinfo {volume} {2}},\ \bibinfo {pages} {e1600093} (\bibinfo {year} {2016})}\BibitemShut {NoStop}%
\bibitem [{\citenamefont {Sosnova}\ \emph {et~al.}(2021)\citenamefont {Sosnova}, \citenamefont {Carter},\ and\ \citenamefont {Monroe}}]{PhysRevA.103.012610}%
  \BibitemOpen
  \bibfield  {author} {\bibinfo {author} {\bibfnamefont {K.}~\bibnamefont {Sosnova}}, \bibinfo {author} {\bibfnamefont {A.}~\bibnamefont {Carter}}, \ and\ \bibinfo {author} {\bibfnamefont {C.}~\bibnamefont {Monroe}},\ }\bibfield  {title} {\enquote {\bibinfo {title} {Character of motional modes for entanglement and sympathetic cooling of mixed-species trapped-ion chains},}\ }\href {\doibase 10.1103/PhysRevA.103.012610} {\bibfield  {journal} {\bibinfo  {journal} {Phys. Rev. A}\ }\textbf {\bibinfo {volume} {103}},\ \bibinfo {pages} {012610} (\bibinfo {year} {2021})}\BibitemShut {NoStop}%
\bibitem [{\citenamefont {Pagano}\ \emph {et~al.}(2018)\citenamefont {Pagano}, \citenamefont {Hess}, \citenamefont {Kaplan}, \citenamefont {Tan}, \citenamefont {Richerme}, \citenamefont {Becker}, \citenamefont {Kyprianidis}, \citenamefont {Zhang}, \citenamefont {Birckelbaw}, \citenamefont {Hernandez}, \citenamefont {Wu},\ and\ \citenamefont {Monroe}}]{Pagano_2019}%
  \BibitemOpen
  \bibfield  {author} {\bibinfo {author} {\bibfnamefont {G}~\bibnamefont {Pagano}}, \bibinfo {author} {\bibfnamefont {P~W}\ \bibnamefont {Hess}}, \bibinfo {author} {\bibfnamefont {H~B}\ \bibnamefont {Kaplan}}, \bibinfo {author} {\bibfnamefont {W~L}\ \bibnamefont {Tan}}, \bibinfo {author} {\bibfnamefont {P}~\bibnamefont {Richerme}}, \bibinfo {author} {\bibfnamefont {P}~\bibnamefont {Becker}}, \bibinfo {author} {\bibfnamefont {A}~\bibnamefont {Kyprianidis}}, \bibinfo {author} {\bibfnamefont {J}~\bibnamefont {Zhang}}, \bibinfo {author} {\bibfnamefont {E}~\bibnamefont {Birckelbaw}}, \bibinfo {author} {\bibfnamefont {M~R}\ \bibnamefont {Hernandez}}, \bibinfo {author} {\bibfnamefont {Y}~\bibnamefont {Wu}}, \ and\ \bibinfo {author} {\bibfnamefont {C}~\bibnamefont {Monroe}},\ }\bibfield  {title} {\enquote {\bibinfo {title} {Cryogenic trapped-ion system for large scale quantum simulation},}\ }\href {\doibase 10.1088/2058-9565/aae0fe} {\bibfield  {journal} {\bibinfo  {journal} {Quantum Science and Technology}\ }\textbf {\bibinfo {volume} {4}},\ \bibinfo {pages} {014004} (\bibinfo {year} {2018})}\BibitemShut {NoStop}%
\bibitem [{\citenamefont {Yang}\ \emph {et~al.}(2022)\citenamefont {Yang}, \citenamefont {Ma}, \citenamefont {Wu}, \citenamefont {Wang}, \citenamefont {Cao}, \citenamefont {Guo}, \citenamefont {Huang}, \citenamefont {Feng}, \citenamefont {Zhou},\ and\ \citenamefont {Duan}}]{yang2022realizing}%
  \BibitemOpen
  \bibfield  {author} {\bibinfo {author} {\bibfnamefont {H.-X.}\ \bibnamefont {Yang}}, \bibinfo {author} {\bibfnamefont {J.-Y.}\ \bibnamefont {Ma}}, \bibinfo {author} {\bibfnamefont {Y.-K.}\ \bibnamefont {Wu}}, \bibinfo {author} {\bibfnamefont {Y.}~\bibnamefont {Wang}}, \bibinfo {author} {\bibfnamefont {M.-M.}\ \bibnamefont {Cao}}, \bibinfo {author} {\bibfnamefont {W.-X.}\ \bibnamefont {Guo}}, \bibinfo {author} {\bibfnamefont {Y.-Y.}\ \bibnamefont {Huang}}, \bibinfo {author} {\bibfnamefont {L.}~\bibnamefont {Feng}}, \bibinfo {author} {\bibfnamefont {Z.-C.}\ \bibnamefont {Zhou}}, \ and\ \bibinfo {author} {\bibfnamefont {L.-M.}\ \bibnamefont {Duan}},\ }\bibfield  {title} {\enquote {\bibinfo {title} {Realizing coherently convertible dual-type qubits with the same ion species},}\ }\href {\doibase 10.1038/s41567-022-01661-5} {\bibfield  {journal} {\bibinfo  {journal} {Nature Physics}\ }\textbf {\bibinfo {volume} {18}},\ \bibinfo {pages} {1058--1061} (\bibinfo {year} {2022})},\ \bibinfo {note} {arXiv:2106.14906}\BibitemShut {NoStop}%
\bibitem [{\citenamefont {Allcock}\ \emph {et~al.}(2021)\citenamefont {Allcock}, \citenamefont {Campbell}, \citenamefont {Chiaverini}, \citenamefont {Chuang}, \citenamefont {Hudson}, \citenamefont {Moore}, \citenamefont {Ransford}, \citenamefont {Roman}, \citenamefont {Sage},\ and\ \citenamefont {Wineland}}]{10.1063/5.0069544}%
  \BibitemOpen
  \bibfield  {author} {\bibinfo {author} {\bibfnamefont {D.~T.~C.}\ \bibnamefont {Allcock}}, \bibinfo {author} {\bibfnamefont {W.~C.}\ \bibnamefont {Campbell}}, \bibinfo {author} {\bibfnamefont {J.}~\bibnamefont {Chiaverini}}, \bibinfo {author} {\bibfnamefont {I.~L.}\ \bibnamefont {Chuang}}, \bibinfo {author} {\bibfnamefont {E.~R.}\ \bibnamefont {Hudson}}, \bibinfo {author} {\bibfnamefont {I.~D.}\ \bibnamefont {Moore}}, \bibinfo {author} {\bibfnamefont {A.}~\bibnamefont {Ransford}}, \bibinfo {author} {\bibfnamefont {C.}~\bibnamefont {Roman}}, \bibinfo {author} {\bibfnamefont {J.~M.}\ \bibnamefont {Sage}}, \ and\ \bibinfo {author} {\bibfnamefont {D.~J.}\ \bibnamefont {Wineland}},\ }\bibfield  {title} {\enquote {\bibinfo {title} {{omg blueprint for trapped ion quantum computing with metastable states}},}\ }\href {\doibase 10.1063/5.0069544} {\bibfield  {journal} {\bibinfo  {journal} {Applied Physics Letters}\ }\textbf {\bibinfo {volume} {119}},\ \bibinfo {pages} {214002} (\bibinfo {year} {2021})},\ \bibinfo {note} {arXiv:2109.01272}\BibitemShut {NoStop}%
\bibitem [{\citenamefont {Feng}\ \emph {et~al.}(2024)\citenamefont {Feng}, \citenamefont {Huang}, \citenamefont {Wu}, \citenamefont {Guo}, \citenamefont {Ma}, \citenamefont {Yang}, \citenamefont {Zhang}, \citenamefont {Wang}, \citenamefont {Huang}, \citenamefont {Zhang}, \citenamefont {Yao}, \citenamefont {Qi}, \citenamefont {Pu}, \citenamefont {Zhou},\ and\ \citenamefont {Duan}}]{Feng2024}%
  \BibitemOpen
  \bibfield  {author} {\bibinfo {author} {\bibfnamefont {L.}~\bibnamefont {Feng}}, \bibinfo {author} {\bibfnamefont {Y.-Y.}\ \bibnamefont {Huang}}, \bibinfo {author} {\bibfnamefont {Y.-K.}\ \bibnamefont {Wu}}, \bibinfo {author} {\bibfnamefont {W.-X.}\ \bibnamefont {Guo}}, \bibinfo {author} {\bibfnamefont {J.-Y.}\ \bibnamefont {Ma}}, \bibinfo {author} {\bibfnamefont {H.-X.}\ \bibnamefont {Yang}}, \bibinfo {author} {\bibfnamefont {L.}~\bibnamefont {Zhang}}, \bibinfo {author} {\bibfnamefont {Y.}~\bibnamefont {Wang}}, \bibinfo {author} {\bibfnamefont {C.-X.}\ \bibnamefont {Huang}}, \bibinfo {author} {\bibfnamefont {C.}~\bibnamefont {Zhang}}, \bibinfo {author} {\bibfnamefont {L.}~\bibnamefont {Yao}}, \bibinfo {author} {\bibfnamefont {B.-X.}\ \bibnamefont {Qi}}, \bibinfo {author} {\bibfnamefont {Y.-F.}\ \bibnamefont {Pu}}, \bibinfo {author} {\bibfnamefont {Z.-C.}\ \bibnamefont {Zhou}}, \ and\ \bibinfo {author} {\bibfnamefont {L.-M.}\ \bibnamefont {Duan}},\ }\bibfield  {title} {\enquote {\bibinfo {title} {Realization of a crosstalk-avoided quantum network node using dual-type qubits of the same ion species},}\ }\href {\doibase 10.1038/s41467-023-44220-z} {\bibfield  {journal} {\bibinfo  {journal} {Nature Communications}\ }\textbf {\bibinfo {volume} {15}},\ \bibinfo {pages} {204} (\bibinfo {year} {2024})}\BibitemShut {NoStop}%
\bibitem [{\citenamefont {Lai}\ \emph {et~al.}(2025)\citenamefont {Lai}, \citenamefont {Wang}, \citenamefont {Shi}, \citenamefont {Cui}, \citenamefont {Wang}, \citenamefont {Zhang}, \citenamefont {Liu}, \citenamefont {Tian}, \citenamefont {Sun}, \citenamefont {Chang}, \citenamefont {Qi}, \citenamefont {Huang}, \citenamefont {Zhou}, \citenamefont {Wu}, \citenamefont {Xu}, \citenamefont {Pu},\ and\ \citenamefont {Duan}}]{PhysRevLett.134.070801}%
  \BibitemOpen
  \bibfield  {author} {\bibinfo {author} {\bibfnamefont {P.-C.}\ \bibnamefont {Lai}}, \bibinfo {author} {\bibfnamefont {Y.}~\bibnamefont {Wang}}, \bibinfo {author} {\bibfnamefont {J.-X.}\ \bibnamefont {Shi}}, \bibinfo {author} {\bibfnamefont {Z.-B.}\ \bibnamefont {Cui}}, \bibinfo {author} {\bibfnamefont {Z.-Q.}\ \bibnamefont {Wang}}, \bibinfo {author} {\bibfnamefont {S.}~\bibnamefont {Zhang}}, \bibinfo {author} {\bibfnamefont {P.-Y.}\ \bibnamefont {Liu}}, \bibinfo {author} {\bibfnamefont {Z.-C.}\ \bibnamefont {Tian}}, \bibinfo {author} {\bibfnamefont {Y.-D.}\ \bibnamefont {Sun}}, \bibinfo {author} {\bibfnamefont {X.-Y.}\ \bibnamefont {Chang}}, \bibinfo {author} {\bibfnamefont {B.-X.}\ \bibnamefont {Qi}}, \bibinfo {author} {\bibfnamefont {Y.-Y.}\ \bibnamefont {Huang}}, \bibinfo {author} {\bibfnamefont {Z.-C.}\ \bibnamefont {Zhou}}, \bibinfo {author} {\bibfnamefont {Y.-K.}\ \bibnamefont {Wu}}, \bibinfo {author} {\bibfnamefont {Y.}~\bibnamefont {Xu}}, \bibinfo {author} {\bibfnamefont {Y.-F.}\ \bibnamefont {Pu}}, \ and\ \bibinfo {author} {\bibfnamefont {L.-M.}\ \bibnamefont {Duan}},\ }\bibfield  {title} {\enquote {\bibinfo {title} {Realization of a crosstalk-free two-ion node for long-distance quantum networking},}\ }\href {\doibase 10.1103/PhysRevLett.134.070801} {\bibfield  {journal} {\bibinfo  {journal} {Phys. Rev. Lett.}\ }\textbf {\bibinfo {volume} {134}},\ \bibinfo {pages} {070801} (\bibinfo {year} {2025})}\BibitemShut {NoStop}%
\bibitem [{\citenamefont {Vizvary}\ \emph {et~al.}(2024)\citenamefont {Vizvary}, \citenamefont {Wall}, \citenamefont {Boguslawski}, \citenamefont {Bareian}, \citenamefont {Derevianko}, \citenamefont {Campbell},\ and\ \citenamefont {Hudson}}]{PhysRevLett.132.263201}%
  \BibitemOpen
  \bibfield  {author} {\bibinfo {author} {\bibfnamefont {Samuel~R.}\ \bibnamefont {Vizvary}}, \bibinfo {author} {\bibfnamefont {Zachary~J.}\ \bibnamefont {Wall}}, \bibinfo {author} {\bibfnamefont {Matthew~J.}\ \bibnamefont {Boguslawski}}, \bibinfo {author} {\bibfnamefont {Michael}\ \bibnamefont {Bareian}}, \bibinfo {author} {\bibfnamefont {Andrei}\ \bibnamefont {Derevianko}}, \bibinfo {author} {\bibfnamefont {Wesley~C.}\ \bibnamefont {Campbell}}, \ and\ \bibinfo {author} {\bibfnamefont {Eric~R.}\ \bibnamefont {Hudson}},\ }\bibfield  {title} {\enquote {\bibinfo {title} {Eliminating qubit-type cross-talk in the $omg$ protocol},}\ }\href {\doibase 10.1103/PhysRevLett.132.263201} {\bibfield  {journal} {\bibinfo  {journal} {Phys. Rev. Lett.}\ }\textbf {\bibinfo {volume} {132}},\ \bibinfo {pages} {263201} (\bibinfo {year} {2024})}\BibitemShut {NoStop}%
\bibitem [{\citenamefont {Shi}\ \emph {et~al.}(2025)\citenamefont {Shi}, \citenamefont {Sinanan-Singh}, \citenamefont {DeBry}, \citenamefont {Todaro}, \citenamefont {Chuang},\ and\ \citenamefont {Chiaverini}}]{PhysRevA.111.L020601}%
  \BibitemOpen
  \bibfield  {author} {\bibinfo {author} {\bibfnamefont {Xiaoyang}\ \bibnamefont {Shi}}, \bibinfo {author} {\bibfnamefont {Jasmine}\ \bibnamefont {Sinanan-Singh}}, \bibinfo {author} {\bibfnamefont {Kyle}\ \bibnamefont {DeBry}}, \bibinfo {author} {\bibfnamefont {Susanna~L.}\ \bibnamefont {Todaro}}, \bibinfo {author} {\bibfnamefont {Isaac~L.}\ \bibnamefont {Chuang}}, \ and\ \bibinfo {author} {\bibfnamefont {John}\ \bibnamefont {Chiaverini}},\ }\bibfield  {title} {\enquote {\bibinfo {title} {Long-lived metastable-qubit memory},}\ }\href {\doibase 10.1103/PhysRevA.111.L020601} {\bibfield  {journal} {\bibinfo  {journal} {Phys. Rev. A}\ }\textbf {\bibinfo {volume} {111}},\ \bibinfo {pages} {L020601} (\bibinfo {year} {2025})}\BibitemShut {NoStop}%
\bibitem [{\citenamefont {Lange}\ \emph {et~al.}(2021)\citenamefont {Lange}, \citenamefont {Peshkov}, \citenamefont {Huntemann}, \citenamefont {Tamm}, \citenamefont {Surzhykov},\ and\ \citenamefont {Peik}}]{PhysRevLett.127.213001}%
  \BibitemOpen
  \bibfield  {author} {\bibinfo {author} {\bibfnamefont {R.}~\bibnamefont {Lange}}, \bibinfo {author} {\bibfnamefont {A.~A.}\ \bibnamefont {Peshkov}}, \bibinfo {author} {\bibfnamefont {N.}~\bibnamefont {Huntemann}}, \bibinfo {author} {\bibfnamefont {Chr.}\ \bibnamefont {Tamm}}, \bibinfo {author} {\bibfnamefont {A.}~\bibnamefont {Surzhykov}}, \ and\ \bibinfo {author} {\bibfnamefont {E.}~\bibnamefont {Peik}},\ }\bibfield  {title} {\enquote {\bibinfo {title} {Lifetime of the $^{2}{F}_{7/2}$ level in ${\mathrm{yb}}^{+}$ for spontaneous emission of electric octupole radiation},}\ }\href {\doibase 10.1103/PhysRevLett.127.213001} {\bibfield  {journal} {\bibinfo  {journal} {Phys. Rev. Lett.}\ }\textbf {\bibinfo {volume} {127}},\ \bibinfo {pages} {213001} (\bibinfo {year} {2021})}\BibitemShut {NoStop}%
\bibitem [{\citenamefont {Duan}\ and\ \citenamefont {Guo}(1997)}]{PhysRevLett.79.1953}%
  \BibitemOpen
  \bibfield  {author} {\bibinfo {author} {\bibfnamefont {Lu-Ming}\ \bibnamefont {Duan}}\ and\ \bibinfo {author} {\bibfnamefont {Guang-Can}\ \bibnamefont {Guo}},\ }\bibfield  {title} {\enquote {\bibinfo {title} {Preserving coherence in quantum computation by pairing quantum bits},}\ }\href {\doibase 10.1103/PhysRevLett.79.1953} {\bibfield  {journal} {\bibinfo  {journal} {Phys. Rev. Lett.}\ }\textbf {\bibinfo {volume} {79}},\ \bibinfo {pages} {1953--1956} (\bibinfo {year} {1997})}\BibitemShut {NoStop}%
\bibitem [{\citenamefont {Lidar}\ \emph {et~al.}(1998)\citenamefont {Lidar}, \citenamefont {Chuang},\ and\ \citenamefont {Whaley}}]{PhysRevLett.81.2594}%
  \BibitemOpen
  \bibfield  {author} {\bibinfo {author} {\bibfnamefont {D.~A.}\ \bibnamefont {Lidar}}, \bibinfo {author} {\bibfnamefont {I.~L.}\ \bibnamefont {Chuang}}, \ and\ \bibinfo {author} {\bibfnamefont {K.~B.}\ \bibnamefont {Whaley}},\ }\bibfield  {title} {\enquote {\bibinfo {title} {Decoherence-free subspaces for quantum computation},}\ }\href {\doibase 10.1103/PhysRevLett.81.2594} {\bibfield  {journal} {\bibinfo  {journal} {Phys. Rev. Lett.}\ }\textbf {\bibinfo {volume} {81}},\ \bibinfo {pages} {2594--2597} (\bibinfo {year} {1998})}\BibitemShut {NoStop}%
\bibitem [{\citenamefont {Kielpinski}\ \emph {et~al.}(2001)\citenamefont {Kielpinski}, \citenamefont {Meyer}, \citenamefont {Rowe}, \citenamefont {Sackett}, \citenamefont {Itano}, \citenamefont {Monroe},\ and\ \citenamefont {Wineland}}]{doi:10.1126/science.1057357}%
  \BibitemOpen
  \bibfield  {author} {\bibinfo {author} {\bibfnamefont {D.}~\bibnamefont {Kielpinski}}, \bibinfo {author} {\bibfnamefont {V.}~\bibnamefont {Meyer}}, \bibinfo {author} {\bibfnamefont {M.~A.}\ \bibnamefont {Rowe}}, \bibinfo {author} {\bibfnamefont {C.~A.}\ \bibnamefont {Sackett}}, \bibinfo {author} {\bibfnamefont {W.~M.}\ \bibnamefont {Itano}}, \bibinfo {author} {\bibfnamefont {C.}~\bibnamefont {Monroe}}, \ and\ \bibinfo {author} {\bibfnamefont {D.~J.}\ \bibnamefont {Wineland}},\ }\bibfield  {title} {\enquote {\bibinfo {title} {A decoherence-free quantum memory using trapped ions},}\ }\href {\doibase 10.1126/science.1057357} {\bibfield  {journal} {\bibinfo  {journal} {Science}\ }\textbf {\bibinfo {volume} {291}},\ \bibinfo {pages} {1013--1015} (\bibinfo {year} {2001})}\BibitemShut {NoStop}%
\bibitem [{\citenamefont {Roos}\ \emph {et~al.}(2004)\citenamefont {Roos}, \citenamefont {Lancaster}, \citenamefont {Riebe}, \citenamefont {H\"affner}, \citenamefont {H\"ansel}, \citenamefont {Gulde}, \citenamefont {Becher}, \citenamefont {Eschner}, \citenamefont {Schmidt-Kaler},\ and\ \citenamefont {Blatt}}]{PhysRevLett.92.220402}%
  \BibitemOpen
  \bibfield  {author} {\bibinfo {author} {\bibfnamefont {C.~F.}\ \bibnamefont {Roos}}, \bibinfo {author} {\bibfnamefont {G.~P.~T.}\ \bibnamefont {Lancaster}}, \bibinfo {author} {\bibfnamefont {M.}~\bibnamefont {Riebe}}, \bibinfo {author} {\bibfnamefont {H.}~\bibnamefont {H\"affner}}, \bibinfo {author} {\bibfnamefont {W.}~\bibnamefont {H\"ansel}}, \bibinfo {author} {\bibfnamefont {S.}~\bibnamefont {Gulde}}, \bibinfo {author} {\bibfnamefont {C.}~\bibnamefont {Becher}}, \bibinfo {author} {\bibfnamefont {J.}~\bibnamefont {Eschner}}, \bibinfo {author} {\bibfnamefont {F.}~\bibnamefont {Schmidt-Kaler}}, \ and\ \bibinfo {author} {\bibfnamefont {R.}~\bibnamefont {Blatt}},\ }\bibfield  {title} {\enquote {\bibinfo {title} {Bell states of atoms with ultralong lifetimes and their tomographic state analysis},}\ }\href {\doibase 10.1103/PhysRevLett.92.220402} {\bibfield  {journal} {\bibinfo  {journal} {Phys. Rev. Lett.}\ }\textbf {\bibinfo {volume} {92}},\ \bibinfo {pages} {220402} (\bibinfo {year} {2004})}\BibitemShut {NoStop}%
\bibitem [{\citenamefont {Buchachenko}\ \emph {et~al.}(2009)\citenamefont {Buchachenko}, \citenamefont {Suleimanov}, \citenamefont {Szcz{\k{e}}{\'s}niak},\ and\ \citenamefont {Cha{\l}asi{\'n}ski}}]{Buchachenko_2009}%
  \BibitemOpen
  \bibfield  {author} {\bibinfo {author} {\bibfnamefont {A~A}\ \bibnamefont {Buchachenko}}, \bibinfo {author} {\bibfnamefont {Yu~V}\ \bibnamefont {Suleimanov}}, \bibinfo {author} {\bibfnamefont {M~M}\ \bibnamefont {Szcz{\k{e}}{\'s}niak}}, \ and\ \bibinfo {author} {\bibfnamefont {G}~\bibnamefont {Cha{\l}asi{\'n}ski}},\ }\bibfield  {title} {\enquote {\bibinfo {title} {Interactions and collisions of cold metal atoms in magnetic traps},}\ }\href {\doibase 10.1088/0031-8949/80/04/048109} {\bibfield  {journal} {\bibinfo  {journal} {Physica Scripta}\ }\textbf {\bibinfo {volume} {80}},\ \bibinfo {pages} {048109} (\bibinfo {year} {2009})}\BibitemShut {NoStop}%
\bibitem [{\citenamefont {Ratschbacher}\ \emph {et~al.}(2013)\citenamefont {Ratschbacher}, \citenamefont {Sias}, \citenamefont {Carcagni}, \citenamefont {Silver}, \citenamefont {Zipkes},\ and\ \citenamefont {K\"ohl}}]{PhysRevLett.110.160402}%
  \BibitemOpen
  \bibfield  {author} {\bibinfo {author} {\bibfnamefont {L.}~\bibnamefont {Ratschbacher}}, \bibinfo {author} {\bibfnamefont {C.}~\bibnamefont {Sias}}, \bibinfo {author} {\bibfnamefont {L.}~\bibnamefont {Carcagni}}, \bibinfo {author} {\bibfnamefont {J.~M.}\ \bibnamefont {Silver}}, \bibinfo {author} {\bibfnamefont {C.}~\bibnamefont {Zipkes}}, \ and\ \bibinfo {author} {\bibfnamefont {M.}~\bibnamefont {K\"ohl}},\ }\bibfield  {title} {\enquote {\bibinfo {title} {Decoherence of a single-ion qubit immersed in a spin-polarized atomic bath},}\ }\href {\doibase 10.1103/PhysRevLett.110.160402} {\bibfield  {journal} {\bibinfo  {journal} {Phys. Rev. Lett.}\ }\textbf {\bibinfo {volume} {110}},\ \bibinfo {pages} {160402} (\bibinfo {year} {2013})}\BibitemShut {NoStop}%
\bibitem [{\citenamefont {Tscherbul}\ \emph {et~al.}(2016)\citenamefont {Tscherbul}, \citenamefont {Brumer},\ and\ \citenamefont {Buchachenko}}]{PhysRevLett.117.143201}%
  \BibitemOpen
  \bibfield  {author} {\bibinfo {author} {\bibfnamefont {Timur~V.}\ \bibnamefont {Tscherbul}}, \bibinfo {author} {\bibfnamefont {Paul}\ \bibnamefont {Brumer}}, \ and\ \bibinfo {author} {\bibfnamefont {Alexei~A.}\ \bibnamefont {Buchachenko}},\ }\bibfield  {title} {\enquote {\bibinfo {title} {Spin-orbit interactions and quantum spin dynamics in cold ion-atom collisions},}\ }\href {\doibase 10.1103/PhysRevLett.117.143201} {\bibfield  {journal} {\bibinfo  {journal} {Phys. Rev. Lett.}\ }\textbf {\bibinfo {volume} {117}},\ \bibinfo {pages} {143201} (\bibinfo {year} {2016})}\BibitemShut {NoStop}%
\bibitem [{\citenamefont {Baldwin}\ \emph {et~al.}(2021)\citenamefont {Baldwin}, \citenamefont {Bjork}, \citenamefont {Foss-Feig}, \citenamefont {Gaebler}, \citenamefont {Hayes}, \citenamefont {Kokish}, \citenamefont {Langer}, \citenamefont {Sedlacek}, \citenamefont {Stack},\ and\ \citenamefont {Vittorini}}]{PhysRevA.103.012603}%
  \BibitemOpen
  \bibfield  {author} {\bibinfo {author} {\bibfnamefont {C.~H.}\ \bibnamefont {Baldwin}}, \bibinfo {author} {\bibfnamefont {B.~J.}\ \bibnamefont {Bjork}}, \bibinfo {author} {\bibfnamefont {M.}~\bibnamefont {Foss-Feig}}, \bibinfo {author} {\bibfnamefont {J.~P.}\ \bibnamefont {Gaebler}}, \bibinfo {author} {\bibfnamefont {D.}~\bibnamefont {Hayes}}, \bibinfo {author} {\bibfnamefont {M.~G.}\ \bibnamefont {Kokish}}, \bibinfo {author} {\bibfnamefont {C.}~\bibnamefont {Langer}}, \bibinfo {author} {\bibfnamefont {J.~A.}\ \bibnamefont {Sedlacek}}, \bibinfo {author} {\bibfnamefont {D.}~\bibnamefont {Stack}}, \ and\ \bibinfo {author} {\bibfnamefont {G.}~\bibnamefont {Vittorini}},\ }\bibfield  {title} {\enquote {\bibinfo {title} {High-fidelity light-shift gate for clock-state qubits},}\ }\href {\doibase 10.1103/PhysRevA.103.012603} {\bibfield  {journal} {\bibinfo  {journal} {Phys. Rev. A}\ }\textbf {\bibinfo {volume} {103}},\ \bibinfo {pages} {012603} (\bibinfo {year} {2021})}\BibitemShut {NoStop}%
\bibitem [{sup()}]{supp}%
  \BibitemOpen
  \href@noop {} {}\bibinfo {note} {See Supplemental Material at URL-will-be-inserted-by-publisher for details about the state preparation, the multi-state detection sequence, the estimation method for the storage lifetime, the leakage model used in numerical simulation, the experimental sequences used in distinguishing different $F_{7/2}$ Zeeman levels and the measured magnetic field gradient at two ions.}\BibitemShut {Stop}%
\bibitem [{\citenamefont {Lin}\ \emph {et~al.}(2013)\citenamefont {Lin}, \citenamefont {Gaebler}, \citenamefont {Reiter}, \citenamefont {Tan}, \citenamefont {Bowler}, \citenamefont {S{\o}rensen}, \citenamefont {Leibfried},\ and\ \citenamefont {Wineland}}]{Lin2013}%
  \BibitemOpen
  \bibfield  {author} {\bibinfo {author} {\bibfnamefont {Y.}~\bibnamefont {Lin}}, \bibinfo {author} {\bibfnamefont {J.~P.}\ \bibnamefont {Gaebler}}, \bibinfo {author} {\bibfnamefont {F.}~\bibnamefont {Reiter}}, \bibinfo {author} {\bibfnamefont {T.~R.}\ \bibnamefont {Tan}}, \bibinfo {author} {\bibfnamefont {R.}~\bibnamefont {Bowler}}, \bibinfo {author} {\bibfnamefont {A.~S.}\ \bibnamefont {S{\o}rensen}}, \bibinfo {author} {\bibfnamefont {D.}~\bibnamefont {Leibfried}}, \ and\ \bibinfo {author} {\bibfnamefont {D.~J.}\ \bibnamefont {Wineland}},\ }\bibfield  {title} {\enquote {\bibinfo {title} {Dissipative production of a maximally entangled steady state of two quantum bits},}\ }\href {\doibase 10.1038/nature12801} {\bibfield  {journal} {\bibinfo  {journal} {Nature}\ }\textbf {\bibinfo {volume} {504}},\ \bibinfo {pages} {415--418} (\bibinfo {year} {2013})}\BibitemShut {NoStop}%
\bibitem [{\citenamefont {Cole}\ \emph {et~al.}(2022)\citenamefont {Cole}, \citenamefont {Erickson}, \citenamefont {Zarantonello}, \citenamefont {Horn}, \citenamefont {Hou}, \citenamefont {Wu}, \citenamefont {Slichter}, \citenamefont {Reiter}, \citenamefont {Koch},\ and\ \citenamefont {Leibfried}}]{PhysRevLett.128.080502}%
  \BibitemOpen
  \bibfield  {author} {\bibinfo {author} {\bibfnamefont {Daniel~C.}\ \bibnamefont {Cole}}, \bibinfo {author} {\bibfnamefont {Stephen~D.}\ \bibnamefont {Erickson}}, \bibinfo {author} {\bibfnamefont {Giorgio}\ \bibnamefont {Zarantonello}}, \bibinfo {author} {\bibfnamefont {Karl~P.}\ \bibnamefont {Horn}}, \bibinfo {author} {\bibfnamefont {Pan-Yu}\ \bibnamefont {Hou}}, \bibinfo {author} {\bibfnamefont {Jenny~J.}\ \bibnamefont {Wu}}, \bibinfo {author} {\bibfnamefont {Daniel~H.}\ \bibnamefont {Slichter}}, \bibinfo {author} {\bibfnamefont {Florentin}\ \bibnamefont {Reiter}}, \bibinfo {author} {\bibfnamefont {Christiane~P.}\ \bibnamefont {Koch}}, \ and\ \bibinfo {author} {\bibfnamefont {Dietrich}\ \bibnamefont {Leibfried}},\ }\bibfield  {title} {\enquote {\bibinfo {title} {Resource-efficient dissipative entanglement of two trapped-ion qubits},}\ }\href {\doibase 10.1103/PhysRevLett.128.080502} {\bibfield  {journal} {\bibinfo  {journal} {Phys. Rev. Lett.}\ }\textbf {\bibinfo {volume} {128}},\ \bibinfo {pages} {080502} (\bibinfo {year} {2022})}\BibitemShut {NoStop}%
\bibitem [{\citenamefont {Grassl}\ \emph {et~al.}(1997)\citenamefont {Grassl}, \citenamefont {Beth},\ and\ \citenamefont {Pellizzari}}]{PhysRevA.56.33}%
  \BibitemOpen
  \bibfield  {author} {\bibinfo {author} {\bibfnamefont {M.}~\bibnamefont {Grassl}}, \bibinfo {author} {\bibfnamefont {Th.}\ \bibnamefont {Beth}}, \ and\ \bibinfo {author} {\bibfnamefont {T.}~\bibnamefont {Pellizzari}},\ }\bibfield  {title} {\enquote {\bibinfo {title} {Codes for the quantum erasure channel},}\ }\href {\doibase 10.1103/PhysRevA.56.33} {\bibfield  {journal} {\bibinfo  {journal} {Phys. Rev. A}\ }\textbf {\bibinfo {volume} {56}},\ \bibinfo {pages} {33--38} (\bibinfo {year} {1997})}\BibitemShut {NoStop}%
\bibitem [{\citenamefont {Bennett}\ \emph {et~al.}(1997)\citenamefont {Bennett}, \citenamefont {DiVincenzo},\ and\ \citenamefont {Smolin}}]{PhysRevLett.78.3217}%
  \BibitemOpen
  \bibfield  {author} {\bibinfo {author} {\bibfnamefont {Charles~H.}\ \bibnamefont {Bennett}}, \bibinfo {author} {\bibfnamefont {David~P.}\ \bibnamefont {DiVincenzo}}, \ and\ \bibinfo {author} {\bibfnamefont {John~A.}\ \bibnamefont {Smolin}},\ }\bibfield  {title} {\enquote {\bibinfo {title} {Capacities of quantum erasure channels},}\ }\href {\doibase 10.1103/PhysRevLett.78.3217} {\bibfield  {journal} {\bibinfo  {journal} {Phys. Rev. Lett.}\ }\textbf {\bibinfo {volume} {78}},\ \bibinfo {pages} {3217--3220} (\bibinfo {year} {1997})}\BibitemShut {NoStop}%
\bibitem [{\citenamefont {Roman}\ \emph {et~al.}(2020)\citenamefont {Roman}, \citenamefont {Ransford}, \citenamefont {Ip},\ and\ \citenamefont {Campbell}}]{Roman2020}%
  \BibitemOpen
  \bibfield  {author} {\bibinfo {author} {\bibfnamefont {Conrad}\ \bibnamefont {Roman}}, \bibinfo {author} {\bibfnamefont {Anthony}\ \bibnamefont {Ransford}}, \bibinfo {author} {\bibfnamefont {Michael}\ \bibnamefont {Ip}}, \ and\ \bibinfo {author} {\bibfnamefont {Wesley~C}\ \bibnamefont {Campbell}},\ }\bibfield  {title} {\enquote {\bibinfo {title} {Coherent control for qubit state readout},}\ }\href {\doibase 10.1088/1367-2630/ab9982} {\bibfield  {journal} {\bibinfo  {journal} {New Journal of Physics}\ }\textbf {\bibinfo {volume} {22}},\ \bibinfo {pages} {073038} (\bibinfo {year} {2020})}\BibitemShut {NoStop}%
\bibitem [{\citenamefont {Edmunds}\ \emph {et~al.}(2021)\citenamefont {Edmunds}, \citenamefont {Tan}, \citenamefont {Milne}, \citenamefont {Singh}, \citenamefont {Biercuk},\ and\ \citenamefont {Hempel}}]{edmunds2021scalable}%
  \BibitemOpen
  \bibfield  {author} {\bibinfo {author} {\bibfnamefont {C.~L.}\ \bibnamefont {Edmunds}}, \bibinfo {author} {\bibfnamefont {T.~R.}\ \bibnamefont {Tan}}, \bibinfo {author} {\bibfnamefont {A.~R.}\ \bibnamefont {Milne}}, \bibinfo {author} {\bibfnamefont {A.}~\bibnamefont {Singh}}, \bibinfo {author} {\bibfnamefont {M.~J.}\ \bibnamefont {Biercuk}}, \ and\ \bibinfo {author} {\bibfnamefont {C.}~\bibnamefont {Hempel}},\ }\bibfield  {title} {\enquote {\bibinfo {title} {Scalable hyperfine qubit state detection via electron shelving in the ${}^{2}{D}_{5/2}$ and ${}^{2}{F}_{7/2}$ manifolds in ${}^{171}{\mathrm{Yb}}^{+}$},}\ }\href {\doibase 10.1103/PhysRevA.104.012606} {\bibfield  {journal} {\bibinfo  {journal} {Phys. Rev. A}\ }\textbf {\bibinfo {volume} {104}},\ \bibinfo {pages} {012606} (\bibinfo {year} {2021})}\BibitemShut {NoStop}%
\bibitem [{\citenamefont {Green}\ and\ \citenamefont {Biercuk}(2015)}]{PhysRevLett.114.120502}%
  \BibitemOpen
  \bibfield  {author} {\bibinfo {author} {\bibfnamefont {Todd~J.}\ \bibnamefont {Green}}\ and\ \bibinfo {author} {\bibfnamefont {Michael~J.}\ \bibnamefont {Biercuk}},\ }\bibfield  {title} {\enquote {\bibinfo {title} {Phase-modulated decoupling and error suppression in qubit-oscillator systems},}\ }\href {\doibase 10.1103/PhysRevLett.114.120502} {\bibfield  {journal} {\bibinfo  {journal} {Phys. Rev. Lett.}\ }\textbf {\bibinfo {volume} {114}},\ \bibinfo {pages} {120502} (\bibinfo {year} {2015})}\BibitemShut {NoStop}%
\bibitem [{\citenamefont {Leung}\ \emph {et~al.}(2018)\citenamefont {Leung}, \citenamefont {Landsman}, \citenamefont {Figgatt}, \citenamefont {Linke}, \citenamefont {Monroe},\ and\ \citenamefont {Brown}}]{PhysRevLett.120.020501}%
  \BibitemOpen
  \bibfield  {author} {\bibinfo {author} {\bibfnamefont {Pak~Hong}\ \bibnamefont {Leung}}, \bibinfo {author} {\bibfnamefont {Kevin~A.}\ \bibnamefont {Landsman}}, \bibinfo {author} {\bibfnamefont {Caroline}\ \bibnamefont {Figgatt}}, \bibinfo {author} {\bibfnamefont {Norbert~M.}\ \bibnamefont {Linke}}, \bibinfo {author} {\bibfnamefont {Christopher}\ \bibnamefont {Monroe}}, \ and\ \bibinfo {author} {\bibfnamefont {Kenneth~R.}\ \bibnamefont {Brown}},\ }\bibfield  {title} {\enquote {\bibinfo {title} {Robust 2-qubit gates in a linear ion crystal using a frequency-modulated driving force},}\ }\href {\doibase 10.1103/PhysRevLett.120.020501} {\bibfield  {journal} {\bibinfo  {journal} {Phys. Rev. Lett.}\ }\textbf {\bibinfo {volume} {120}},\ \bibinfo {pages} {020501} (\bibinfo {year} {2018})}\BibitemShut {NoStop}%
\bibitem [{\citenamefont {Tan}\ \emph {et~al.}(2021)\citenamefont {Tan}, \citenamefont {Edmunds}, \citenamefont {Milne}, \citenamefont {Biercuk},\ and\ \citenamefont {Hempel}}]{tan2021QZ_coeff}%
  \BibitemOpen
  \bibfield  {author} {\bibinfo {author} {\bibfnamefont {TR}~\bibnamefont {Tan}}, \bibinfo {author} {\bibfnamefont {CL}~\bibnamefont {Edmunds}}, \bibinfo {author} {\bibfnamefont {AR}~\bibnamefont {Milne}}, \bibinfo {author} {\bibfnamefont {MJ}~\bibnamefont {Biercuk}}, \ and\ \bibinfo {author} {\bibfnamefont {C}~\bibnamefont {Hempel}},\ }\bibfield  {title} {\enquote {\bibinfo {title} {Precision characterization of the ${}^{2}{D}_{5/2}$ state and the quadratic zeeman coefficient in ${}^{171}{\mathrm{Yb}}^{+}$},}\ }\href@noop {} {\bibfield  {journal} {\bibinfo  {journal} {Physical Review A}\ }\textbf {\bibinfo {volume} {104}},\ \bibinfo {pages} {L010802} (\bibinfo {year} {2021})}\BibitemShut {NoStop}%
\bibitem [{\citenamefont {Vanier}\ and\ \citenamefont {Audoin}(1989)}]{Vanier1989book}%
  \BibitemOpen
  \bibfield  {author} {\bibinfo {author} {\bibfnamefont {J}~\bibnamefont {Vanier}}\ and\ \bibinfo {author} {\bibfnamefont {C}~\bibnamefont {Audoin}},\ }\href@noop {} {\emph {\bibinfo {title} {The Quantum Physics of Atomic Frequency Standards (1st ed.)}}}\ (\bibinfo  {publisher} {CRC Press},\ \bibinfo {year} {1989})\BibitemShut {NoStop}%
\end{thebibliography}
%

\end{document}


\title{Supplemental Material for
``Long-time storage of a decoherence-free subspace logical qubit in a dual-type quantum memory''}

\author{Y.-L. Xu}
\thanks{These authors contribute equally to this work}%
\affiliation{Center for Quantum Information, Institute for Interdisciplinary Information Sciences, Tsinghua University, Beijing 100084, PR China}

\author{L. Zhang}
\thanks{These authors contribute equally to this work}%
\affiliation{Center for Quantum Information, Institute for Interdisciplinary Information Sciences, Tsinghua University, Beijing 100084, PR China}

\author{C. Zhang}
\thanks{These authors contribute equally to this work}%
\affiliation{HYQ Co., Ltd., Beijing 100176, P. R. China}

\author{Y.-K. Wu}

\affiliation{Center for Quantum Information, Institute for Interdisciplinary Information Sciences, Tsinghua University, Beijing 100084, PR China}
\affiliation{Hefei National Laboratory, Hefei 230088, PR China}

\author{Y.-Y. Chen}
\affiliation{Center for Quantum Information, Institute for Interdisciplinary Information Sciences, Tsinghua University, Beijing 100084, PR China}

\author{C.-X. Huang}
\affiliation{Center for Quantum Information, Institute for Interdisciplinary Information Sciences, Tsinghua University, Beijing 100084, PR China}

\author{Z.-B. Cui}
\affiliation{Center for Quantum Information, Institute for Interdisciplinary Information Sciences, Tsinghua University, Beijing 100084, PR China}

\author{R. Yao}
\affiliation{HYQ Co., Ltd., Beijing 100176, P. R. China}

\author{W.-Q. Lian}
\affiliation{HYQ Co., Ltd., Beijing 100176, P. R. China}

\author{J.-Y. Ma}
\affiliation{HYQ Co., Ltd., Beijing 100176, P. R. China}

\author{W.-X. Guo}
\affiliation{HYQ Co., Ltd., Beijing 100176, P. R. China}

\author{B.-X. Qi}
\affiliation{Center for Quantum Information, Institute for Interdisciplinary Information Sciences, Tsinghua University, Beijing 100084, PR China}

\author{P.-Y. Hou}
\affiliation{Center for Quantum Information, Institute for Interdisciplinary Information Sciences, Tsinghua University, Beijing 100084, PR China}
\affiliation{Hefei National Laboratory, Hefei 230088, PR China}

\author{Y.-F. Pu}
\affiliation{Center for Quantum Information, Institute for Interdisciplinary Information Sciences, Tsinghua University, Beijing 100084, PR China}
\affiliation{Hefei National Laboratory, Hefei 230088, PR China}

\author{Z.-C. Zhou}
\affiliation{Center for Quantum Information, Institute for Interdisciplinary Information Sciences, Tsinghua University, Beijing 100084, PR China}
\affiliation{Hefei National Laboratory, Hefei 230088, PR China}

\author{L. He}
\affiliation{Center for Quantum Information, Institute for Interdisciplinary Information Sciences, Tsinghua University, Beijing 100084, PR China}
\affiliation{Hefei National Laboratory, Hefei 230088, PR China}

\author{L.-M. Duan}
\email{lmduan@tsinghua.edu.cn}
\affiliation{Center for Quantum Information, Institute for Interdisciplinary Information Sciences, Tsinghua University, Beijing 100084, PR China}
\affiliation{Hefei National Laboratory, Hefei 230088, PR China}

\maketitle

\makeatletter
\renewcommand{\thefigure}{S\arabic{figure}}
\renewcommand{\thetable}{S\arabic{table}}
\renewcommand{\theequation}{S\arabic{equation}}
\makeatother

\section{Preparation of DFS logical states and coherent conversion}
\begin{figure}[h]
    \centering
    \includegraphics{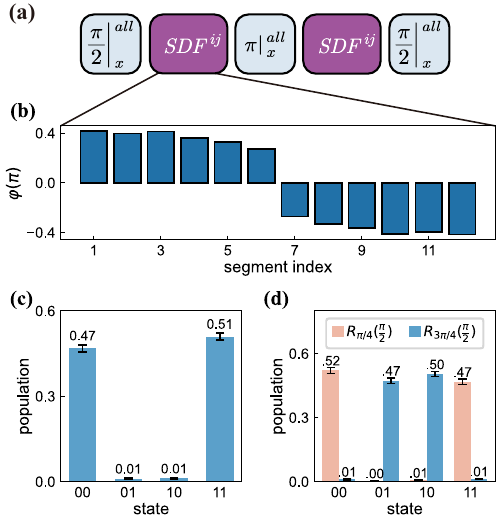}
    \caption{(a) Experimental sequence to prepare a maximally entangled state using global microwave pulses and spin-dependent forces (SDF) induced by counter-propagating $411\,$nm laser beams. (b) Phase modulation sequence for the entangling gate. (c) Population distribution of the prepared maximally entangled state. (d) Parity of the prepared maximally entangled state after a global $\pi/2$ pulse with a phase of $\pi/4$ (red) or $3\pi/4$ (blue).}
    \label{fig:S1}
\end{figure}
At the beginning of each experimental sequence, we first perform Doppler cooling and resolved sideband cooling to initialize the motional state of the ions close to the ground state. Then we apply an optical pumping pulse to prepare the ions in $|0_S\rangle$. To further prepare the DFS logical states $|0_S 1_S\rangle$ and $|1_S 0_S\rangle$ in the $S$-type, we use a focused $411\,$nm laser beam, red detuned by $2\pi\times 26\,$MHz from the $|0_S\rangle \leftrightarrow |D_{5/2},F=2,m_F=-2\rangle$ transition, to generate a $\pi$ phase shift on a selected target ion by the AC Stark shift. When sandwiched by two global microwave $\pi/2$ pulses, this results in the selective flip of any target ion. We perform electron shelving \cite{yang2022realizing,Roman2020,edmunds2021scalable} to measure the prepared computational basis states, and obtain a SPAM error of about $0.7(3)\%$.

To prepare superposition states in the DFS like $(|0_S 1_S\rangle\pm|1_S 0_S\rangle)/\sqrt{2}$, we apply a two-qubit entangling gate on the two memory ions. Specifically, we use counter-propagating $411\,$nm laser beams, again red-detuned from the $|0_S\rangle \leftrightarrow |D_{5/2},F=2,m_F=-2\rangle$ transition, to realize a light shift gate \cite{PhysRevA.103.012603}. This ion-laser interaction generates a spin-dependent force (SDF) on the $|0_S\rangle$ level. To prepare a maximally entangled state, we execute two SDF pulses in the middle of global microwave $\pi/2$, $\pi$ and $\pi/2$ pulses along the $X$ direction as shown in Fig.~\ref{fig:S1}(a). Since we have three ions, the SDF pulse sequence needs to decouple all the three transverse collective phonon modes at the frequencies of $2\pi\times (1.298,\,1.347,\,1.381)\,$MHz. Here we set the beat note frequency of $411\,$nm laser to $\mu=2\pi\times 1.396\,$MHz, and use a phase modulation sequence \cite{PhysRevLett.114.120502,PhysRevLett.120.020501} with $12$ anti-symmetric segments and a total duration of $150\,\upmu$s for each SDF pulse. 
The anti-symmetric phase modulation sequence is numerically optimized to be $\varphi=\pi\times (0.417, 0.398, 0.415, 0.363, 0.331, 0.271, -0.271, -0.331, -0.363, -0.415, -0.398, -0.417)$ as visualized in Fig.~\ref{fig:S1}(b). We further include a $\mathrm{sin}^2$-shaped amplitude ramping with a rise/fall time of $2\,\upmu$s at the beginning/end of each segment to suppress the off-resonant excitation to the $D_{5/2}$ levels. Ideally, this pulse sequence generates a maximally entangled state $(|00\rangle - i|11\rangle)/\sqrt{2}$. We measure its population in Fig.~\ref{fig:S1}(c) and its parity oscillation in Fig.~\ref{fig:S1}(d), and obtain an entanglement fidelity of $97.2(3)\%$. Note that, here instead of scanning the phase $\phi$ of an analysis $R_{\phi}(\pi/2)$ pulse and fitting the parity contrast, we directly measure the parity at the ideal maximum ($\phi=\pi/4$) and the minimum ($\phi=3\pi/4$) to ensure that the obtained fidelity correctly represents the desired entangled state with the phase of $-i$. 
To further turn this state into $(|0_S 1_S\rangle+|1_S 0_S\rangle)/\sqrt{2}$, we apply a global microwave $\pi/2$ rotation along the $(\hat{x}-\hat{y})/\sqrt{2}$ axis where $\hat{x}$ and $\hat{y}$ are the unit vectors along the $X$ and $Y$ directions, respectively. Further turning this state into $(|0_S 1_S\rangle-|1_S 0_S\rangle)/\sqrt{2}$ just requires an additional single-qubit $\pi$ phase shift.

As shown by the energy levels in Fig.~1(b), once encoded into the DFS, the two memory qubits can be converted into the $F$-type spanned by $|0_F\rangle \equiv |F_{7/2},F=3,m_F=0\rangle$ and $|1_F\rangle \equiv |F_{7/2},F=4,m_F=0\rangle$ using global bichromatic $411\,$nm and $3432\,$nm laser $\pi$ pulses via the intermediate $|0_D\rangle \equiv |D_{5/2},F=2,m_F=0\rangle$ and $|1_D\rangle \equiv |D_{5/2},F=3,m_F=0\rangle$ states. To prevent the $S$-type coolant ion from being affected by this global conversion pulse, we first shelve it into $|D_{5/2},F=2,m_F=-1\rangle$ using a focused $411\,$nm $\pi$ pulse. After the conversion of the memory ions, we pump the coolant ion back to the $S_{1/2}$ manifolds by a global $976\,$nm repump laser. This laser also turns any residual population of the two memory ions in $D_{5/2}$ due to imperfect qubit type conversion back into $S_{1/2}$. Therefore, after this step we apply a $370\,$nm laser beam for $1\,$ms to illuminate the ions in the $S_{1/2}$ manifold and collect the scattered photons by an electron-multiplying charge-coupled device (EMCCD) camera [indicated by the camera symbol in the experimental sequence in Fig.~1(c)]. For a successful preparation of an $F$-type DFS logical state, both memory ions should appear dark in the image. Otherwise, if any memory ion is detected to be bright, we simply discard this trial and initialize the state again.

\section{Maximum likelihood estimation for the storage lifetime}
We use the maximum likelihood method to fit the coherence time of the stored quantum states. For the DFS logical states, we use an exponential decay model $F(T;A,\tau)=(1+A e^{-T/\tau})/2$ because these states are immune to the low-frequency phase noises. As for the physical states or the non-DFS logical states, we use a Gaussian decay model $F(T;A,\tau)=(1+A e^{-T^2/\tau^2})/2$. Given the theoretical model $F(T;A,\tau)$, and the experimental data, i.e. the measured average fidelity $F_i$ over $R_i$ repetitions at the storage time $T_i$ ($i=1,\,\cdots,\,K$), we compute the log likelihood function
\begin{align}
L(A,\tau) = \sum_{i=1}^K \big\{& \ln C(R_i, F_i R_i) +  F_i R_i \ln F(T_i;A,\tau) \nonumber\\
&+ (1-F_i)R_i \ln [1-F(T_i;A,\tau)] \big\},
\end{align}
where $C(\cdot, \cdot)$ is the binomial coefficient and is independent of the fitting parameters $A$ and $\tau$. We numerically maximize $L(A,\tau)$ to obtain the best estimator $\hat{A}$ and $\hat{\tau}$ with respect to the experimental data.

To further estimate the confidence interval, we perform Monte Carlo sampling to generate simulated data at the experimentally chosen time points $T_i$ with $R_i$ repetitions ($i=1,\,\cdots,\,K$), again assuming a binomial distribution with a success probability of $F(T_i;\hat{A},\hat{\tau})$. With these simulated fidelities $F_i^\prime$, we can apply the maximum likelihood method again to compute the corresponding fitting parameters $\hat{A}^\prime$ and $\hat{\tau}^\prime$. We can repeat this sampling process to obtain a joint distribution for the fitting parameters $A$ and $\tau$.

In Fig.~3 of the main text, we use the joint distribution of $A$ and $\tau$ to get the confidence interval of the fitting curve, namely the confidence interval of predicting the state fidelity $F(T)$ at each chosen storage time $T$. As for a confidence interval of the coherence time $\tau$ alone, we integrate out the variable $A$ and obtain the marginal distribution, which allows us to estimate an asymmetric confidence interval for $\tau$.

\section{Numerical simulation}
In Fig.~2, we use a Lindblad master equation to simulate the dynamics of two ions
\begin{equation}
\frac{d\rho}{dt} = \sum_{ik} \left(L_{ik} \rho L_{ik}^\dag - \frac{1}{2}L_{ik}^\dag L_{ik} \rho - \frac{1}{2} \rho L_{ik}^\dag L_{ik}\right),
\end{equation}
where the index $i=1,\,2$ represents the two memory ions, and $k$ denotes a group of indices for different decoherence terms on each ion.

Specifically, we consider the leakage error between all the neighboring Zeeman levels in the $F_{7/2}$ manifold with a same leakage rate $\gamma$
\begin{equation}
L_{i;F,m_F,+} = \sqrt{\gamma} |F_{7/2},F,m_F+1\rangle_i\langle F_{7/2},F,m_F|,
\end{equation}
and
\begin{equation}
L_{i;F,m_F,-} = \sqrt{\gamma} |F_{7/2},F,m_F\rangle_i\langle F_{7/2},F,m_F+1|,
\end{equation}
where $F=3,\,4$, and $m_F=-F,\,-F+1,\,\cdots,\,F-1$.

Also we include a dephasing term for each memory qubit with a dephasing rate of $\gamma_d$
\begin{equation}
L_{i;d} = \sqrt{\gamma_d} (|0_F\rangle_i\langle 0_F|-|1_F\rangle_i\langle 1_F|).
\end{equation}
In principle the phase fluctuation exists between all the pairs of levels, but because we are only interested in the qubit levels $|0_F\rangle \equiv |F_{7/2},3,0\rangle$ and $|1_F\rangle \equiv |F_{7/2},4,0\rangle$, and are treating the population in all the other levels as leakage errors, their dephasing errors are not relevant to our results. Also note that here we assume independent dephasing terms for the two memory qubits. This is because the collective dephasing error over the two ions does not affect the DFS logical qubit.

Given the above error model, we fit the leakage rate $\gamma$ from the measured leakage data (about $12\%$ leakage after $T=800\,$s storage), and fit the dephasing rate $\gamma_d$ from the post-selected leakage-free data. As we can see from Fig.~2, the leakage rate is much larger than the residual dephasing rate, making it necessary to distinguish such leakage events as we describe in the main text.

\section{Multi-state detection sequence and calibration of multi-state detection errors}
\begin{figure}[h]
    \centering
    \includegraphics[width=\linewidth]{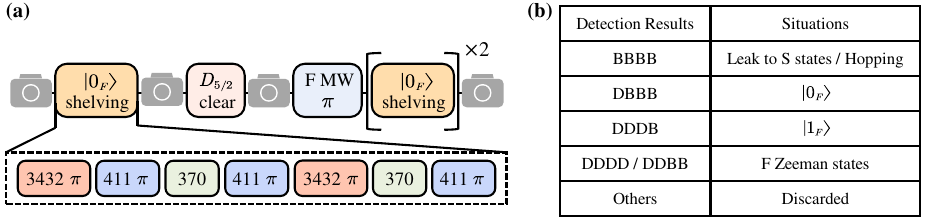}
    \caption{(a) Experimental sequence for multi-state detection of the memory ions. We use bichromatic $3432\,$nm $\pi$ pulses, monochromatic $411\,$nm $\pi$ pulses and $370\,$nm optical pumping pulses to shelve the $|0_F\rangle$ state to $S_{1/2}$. Combined with a global $976\,$nm repump pulse and a microwave $\pi$ pulse, we can distinguish $|0_F\rangle$, $|1_F\rangle$ and the leaked states from the outcomes of multiple detection stages. (b) We interpret the four sequential detection outcomes (``D'' for dark and ``B'' for bright) in (a) into $|0_F\rangle$, $|1_F\rangle$ or the leaked states of the memory ions.}
    \label{fig:S2}
\end{figure}
Experimental sequence in Fig.~\ref{fig:S2}(a) where the four binary detections (bright or dark for each ion) give us sufficient information to distinguish the $|0_F\rangle$ and $|1_F\rangle$ states from leakage or ion-hopping errors during the storage according to the lookup table in Fig.~\ref{fig:S2}(b). Since the memory ions are encoded in the $F$-type which is dark under the $370\,$nm detection laser, a bright ion in the first detection stage indicates a leakage to $S_{1/2}$ or a position exchange with the coolant ion due to the ion hopping, as shown by the first syndrome in Fig.~\ref{fig:S2}(b). Then we use a sequence of global bichromatic $3432\,$nm laser (resonant to $|0_F\rangle \leftrightarrow |0_D\rangle$ and $|1_F\rangle \leftrightarrow |1_D\rangle$), monochromatic $411\,$nm laser (resonant to $|0_D\rangle \leftrightarrow |0_S\rangle$) and $370\,$nm optical pumping laser to selectively shelve the $|0_F\rangle$ state back to the $|S_{1/2},F=1\rangle$ levels, as shown by the $|0_F\rangle$ shelving sequence in the dashed box in Fig.~\ref{fig:S2}(a). In practice, these $\pi$ pulses are not perfect, so we apply another $411\,$nm $\pi$ pulse to suppress the residual population in $|0_D\rangle$, followed by another $3432\,$nm $\pi$ pulse for that in $|0_F\rangle$. The latter $3432\,$nm pulse also serves to restore the population of $|1_F\rangle$, which has previously been transferred into $|1_D\rangle$. Finally we use an additional $370\,$nm optical pumping pulse and a $411\,$nm $\pi$ pulse to clear the newly generated population in $|0_D\rangle$. This explains the second situation in Fig.~\ref{fig:S2}(b) where for the $|0_F\rangle$ state we expect the ion to be bright after the second detection stage. On the other hand, if the ion is originally in $|1_F\rangle$, it will appear dark until we apply a microwave $\pi$ pulse to flip it into $|0_F\rangle$ and further transfer it into $S_{1/2}$ by the $|0_F\rangle$ shelving, which is the third case in Fig.~\ref{fig:S2}(b). Here we apply two $|0_F\rangle$ shelving sequences to suppress the imperfection in the $3432\,$nm $\pi$ pulse.

Because the Zeeman splitting of $9.41\,$MHz for the $|F_{7/2},F=3\rangle$ levels is close to that of $10.24\,$MHz for the $|D_{5/2},F=2\rangle$ levels under our magnetic field of $B=5.23\,$G, during the $|0_F\rangle$ shelving sequence, leakage errors to the $|F_{7/2},F=3,m_F=\pm 1\rangle$ Zeeman levels have a small probability to be converted into $|D_{5/2},F=2,m_F=\pm 1\rangle$ levels by the $3432\,$nm $\pi$ pulse with a Rabi rate of about $200\,$kHz. Such a residual population in $D_{5/2}$ can be identified by a $976\,$nm repump laser [$D_{5/2}$ clear in Fig.~\ref{fig:S2}(a)] and a subsequent detection stage. Otherwise, if the leakage population remains in $F_{7/2}$ Zeeman levels, the ion will stay dark for all the four detection events. These two situations correspond to the fourth syndrome in Fig.~\ref{fig:S2}(b). Finally, the other detection patterns may be explained by some higher-order error events and will be discarded. In practice, we find the probability for such events to be below $0.2\%$. 
We summarize the calibration results for the multi-state detection errors in Table~\ref{tab:S1}, from which we can find that the fidelity for the $|1_F\rangle$ is lower at $93.4\%$. This is mainly due to the imperfect $3432\,$nm $\pi$ pulse, which results in a residual population in $|1_D\rangle$ after the $|0_F\rangle$-shelving sequence. The residual population is then misidentified as a leakage to the $F_{7/2}$ Zeeman levels with bright events detected in the third detection stage.
\begin{table}[h]
\caption{\label{tab:S1}
Probability distribution of multi-state detection outcomes for different input states. The notation $F(a, b)$ represents the state $|F_{7/2},F=a,m_F=b\rangle$. Less than $0.2\%$ detection events, which cannot be explained by the patterns in Fig.~1(e) of the main text, and may come from high-order error events, have been discarded.}
\begin{ruledtabular}
\begin{tabular}{cccc}
Input state & $P_{|0_F\rangle}(\%)$ & $P_{|1_F\rangle}(\%)$ & $P_{\mathrm{Zeeman}}(\%)$\\
\hline
$|0_F\rangle$ & $99.0(2)$ & $0.05(5)$ & $0.95(0.22)$\\
$|1_F\rangle$ & $0.3(1)$ & $93.4(6)$ & $6.3(6)$\\
$F(3,+1)$ & $0.2(1)$ & $0.1(1)$ & $99.7(1)$ \\
$F(3,-1)$ & $0.4(1)$ & $0.2(1)$ & $99.4(2)$\\
$F(4,+1)$ & $0.2(1)$ & $0.6(1)$ & $99.2(1)$\\
$F(4,-1)$ & $0.2(1)$ & $0.7(1)$ & $99.1(1)$\\
\end{tabular}
\end{ruledtabular}
\end{table}
\section{Magnetic field gradient at memory ions}
\begin{figure}[ht]
    \centering
    \includegraphics{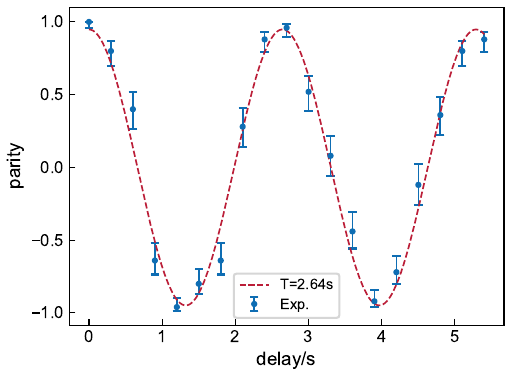}
    \caption{Parity oscillation over the storage time $T$ for an initial $|+^\prime_L\rangle\equiv (|0^\prime 1^\prime\rangle + |1^\prime 0^\prime\rangle)/\sqrt{2}$ state which is sensitive to the magnetic field. The error bars represent $68\%$ confidence interval. An oscillation period of $T=2.64(2)\,$s is fitted.}
    \label{fig:S3}
\end{figure}

Ideally, the DFS logical states are immune to global phase noise on both memory ions and there is no need for a long dynamical decoupling sequence. However, a small magnetic field gradient at the positions of the two ions can result in their frequency difference, and hence the oscillation between the DFS logical states $|\pm_L\rangle\equiv (|0_F 1_F\rangle \pm |1_F 0_F\rangle)/\sqrt{2}$ during the storage. Note that, although the $|0_F\rangle$ and $|1_F\rangle$ states form a pair of clock transition whose frequency is first-order insensitive to the magnetic field, for the long-time storage we need to consider the second-order sensitivity of $354\,\mathrm{Hz}/\mathrm{G}^2$ \cite{tan2021QZ_coeff,Vanier1989book}.

Here to calibrate this nonzero magnetic field gradient, we amplify the sensitivity to the magnetic field by encoding into the Zeeman levels $|0^\prime\rangle \equiv |S_{1/2},F=0,m_F=0\rangle$ and $|1^\prime\rangle\equiv |S_{1/2},F=1,m_F=1\rangle$. Then we measure the oscillation between $|\pm^\prime_L\rangle\equiv (|0^\prime 1^\prime\rangle \pm |1^\prime 0^\prime\rangle)/\sqrt{2}$ in Fig.~\ref{fig:S3}. As we show in the main text, ideally the logical state $|+^\prime\rangle$ has an even parity after a global $\pi/2$ pulse, while the logical state $|-^\prime\rangle$ has an odd parity. From the observed oscillation, we deduce a magnetic field difference of $2.70(2)\times 10^{-7}\,$G between the two memory ions or a local gradient of $416(3)\,\mathrm{G}/\mathrm{m}$. This value corresponds to about $1.0\times 10^{-3}\,$Hz frequency difference for the $F$-type qubits encoded in the two memory ions, and can become non-negligible as the storage time goes above $1000\,$s.
In the main text, we insert two spin echoes into the storage sequence to suppress this oscillation due to a fixed magnetic-field gradient.

\section{Distinguish different $F_{7/2}$ Zeeman levels}
\begin{figure}[ht]
    \centering
    \includegraphics[scale=0.98]{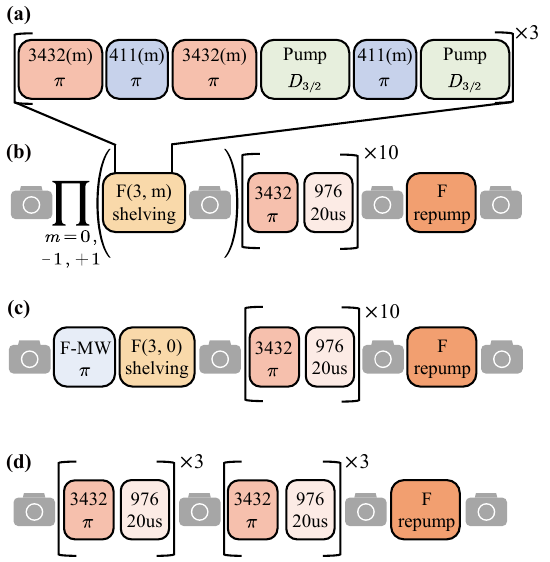}
    \caption{Different leakage error detection sequences for different purposes. (a) The $F(3,m)$-shelving sequence used to transfer the population from $|F_{7/2},F=3,m_F=m\rangle$ to $S_{1/2}$ and $D_{3/2}$. (b) The sequence used to distinguish the leaked states from the initial state $|F_{7/2}, F=3, m_F=0\rangle$. (c) The sequence used to distinguish the leaked states from the initial state $|F_{7/2},F=4,m_F=0\rangle$. (d) The sequence to measure the dependence of the leakage rate versus the temperature.}
    \label{fig:S4}
\end{figure}

In Fig.~1(d) of the main text, we describe the experimental sequence to distinguish the leakage error from the measured qubit state $|0_F\rangle$ and $|1_F\rangle$. To further examine the population on different leaked Zeeman levels, we use the sequence shown in Fig.~\ref{fig:S4}. Similar to the $|0_F\rangle$ shelving sequence in Fig.~1(d) of the main text, here we use a $F(3,m)$-shelving sequence to selectively transfer the population in $|F_{7/2},F=3,m_F=m\rangle$ to the $S_{1/2}$ and $D_{3/2}$ levels which are bright under the $370\,$nm detection laser and the $935\,$nm repump laser. As shown in Fig.~\ref{fig:S4}(a), this sequence consists of a bichromatic $3432\,$nm $\pi$ pulse and a monochromatic $411\,$nm $\pi$ pulse to convert the population in $|F_{7/2},F=3,m_F=m\rangle$ to $|S_{1/2},F=0,m_F=0\rangle$ via the intermediate $|D_{5/2},F=2,m_F=m\rangle$. In this experiment we focus on $m=0,\pm 1$ because they already account for the zeroth and the first order leakage error. Then we apply a second bichromatic $3432\,$nm $\pi$ pulse to restore the population of the $|F_{7/2},F=4\rangle$ levels, which has been transferred to $|D_{5/2},F=3\rangle$ levels temporarily. To support multiple sheving attempts or the following shelving of other Zeeman levels, we depopulate the $S_{1/2}$ levels by a $370\,$nm beam without turning on the $935\,$nm repump beam, such that all the population in $S_{1/2}$ can be converted into $D_{3/2}$. Finally, we apply another monochromatic $411\,$nm $\pi$ pulse to suppress the residual population on $|D_{5/2},F=2,m_F=m\rangle$ due to the pulse imperfection, and again move the population to $D_{3/2}$ through the $370\,$nm optical pumping beam. Overall we repeat this sequence for three times to further suppress the pulse imperfection.

With the $F(3,m)$ shelving sequence as a tool, now we can use the sequence in Fig.~\ref{fig:S4}(b) to diagnose the leaked population from an initial state $|F_{7/2}, F=3, m_F=0\rangle$. We use a first $370\,$nm detection beam with the $935\,$nm repump beam turned on (indicated by the camera symbol) to detect the leakage error to the $S_{1/2}$ states. Then we sequentially perform the $F(3,m)$ shelving sequence for $m=0,-1,+1$, each followed by a detection stage, to distinguish the population in $|F_{7/2}, F=3, m_F=0,\pm 1\rangle$ which constitutes the majority of the final population. For the population in $|F_{7/2},F=4,m_F=0,\pm 1\rangle$, we apply ten cycles of $3432\,$nm $\pi$ pulses and $976\,$nm repump pulses for the $|D_{5/2},F=3\rangle$ levels to clear their population and move them all to the $S_{1/2}$ levels for detection. Finally, we use a $5\,$ms $F$ repump pulse, which consists of a continuous $3432\,$nm laser and a $976\,$nm repump laser, to depopulate all the $F_{7/2}$ Zeeman levels with $-2\le m_F\le 2$. Note that $|F_{7/2},F=3,m_F=\pm 3\rangle$ may not be efficiently repumped because of the lack of corresponding Zeeman levels in the $D_{5/2}$ manifold. Nevertheless, such an ``all dark'' detection event has a vanishingly small probability in our experiment as shown in Fig.~4(a) of the main text, suggesting that the leakage error mainly occurs between the neighboring Zeeman levels.

To diagnose the leaked population from an initial state $|F_{7/2}, F=4, m_F=0\rangle$, we use the experimental sequence in Fig.~\ref{fig:S4}(c). Here because the Zeeman splitting for the $|F_{7/2},F=4\rangle$, $|D_{5/2},F=3\rangle$ and $|S_{1/2},F=1\rangle$ levels are identical, it is difficult to selectively transfer the population of $|F_{7/2},F=4\rangle$ Zeeman levels directly. Instead, we apply a microwave $\pi$ pulse to exchange the population of $|F_{7/2}, F=4, m_F=0\rangle$ and $|F_{7/2}, F=3, m_F=0\rangle$, and then use the $F(3,0)$ shelving sequence to detect its population. Then again we use ten cycles of $3432\,$nm $\pi$ pulses and $976\,$nm repump pulses to move the population in $|F_{7/2},F=4,m_F=\pm1,\pm2,\pm3\rangle$, as well as the original population in $|F_{7/2}, F=3, m_F=0\rangle$ which is now exchanged into $|F_{7/2}, F=4, m_F=0\rangle$, back to $S_{1/2}$ for detection. Finally, we perform the $F$ repump pulse for remaining population in $|F_{7/2},F=3,m_F=\pm 1,\pm2\rangle$.

We utilize the multi-state detection sequences mentioned above to search for the source that leads to the Zeeman leakage. By tuning different experimental conditions, we find this leakage rate to be independent of the laser power, the shielding of the external magnetic field noise, the compensation of the micromotion of the ions or the RF power input to the trap. Instead, the leakage rate increases when the vacuum condition of our system degrades, suggesting it may originate from the collision of the ions with the background gas molecules. 

Finally, when studying the dependence of the leakage rate on the temperature, we are not interested in distinguishing different Zeeman levels. Also, based on the measurement results in Fig.~4(a) of the main text, we find that the leakage rates for the $|F_{7/2}, F=3, m_F=0\rangle$ and $|F_{7/2}, F=4, m_F=0\rangle$ levels are identical within the experimental error bars. Therefore, we initialize the ions in $|F_{7/2}, F=3, m_F=0\rangle$, and execute a simpler sequence to obtain the leakage rate, as shown in Fig.~\ref{fig:S4}(d). Specifically, rather than using a $411\,$nm $\pi$ pulse, we use a continuous $976\,$nm repump beam for the $|D_{5/2},F=2\rangle$ levels, together with the $3432\,$nm $\pi$ pulse, to selectively convert the population in $|F_{7/2}, F=3, m_F=0\rangle$ into $S_{1/2}$. We repeat this sequence for three times to suppress the pulse error. Then after the detection stage, we execute this sequence for another three cycles to verify the vanishing residual population on $|F_{7/2}, F=3, m_F=0\rangle$. Therefore, we attribute the remaining dark events as the leakage error, which can be brought back to $S_{1/2}$ by the $F$ repump pulse. Note that, as we describe in the main text, the $|F_{7/2},F=3\rangle$ states have a Zeeman splitting of $9.41\,$MHz while the $|D_{5/2},F=2\rangle$ levels have a Zeeman splitting of $10.24\,$MHz under the $B=5.23\,$G magnetic field. Therefore there is a small probability for the leaked population in $|F_{7/2}, F=3, m_F=\pm 1\rangle$ to be transferred into $|D_{5/2}, F=2, m_F=\pm 1\rangle$ by the $3432\,$nm $\pi$ pulse. Experimentally we measure this probability to be about $5\%$ after the six pulses, and decide that it will not influence our study of the scaling of the leakage error.

%